\documentclass[pre,showpacs,showkeys,preprintnumbers,amsmath,amssymb,superscriptaddress,twocolumn]{revtex4-2}
\usepackage{graphicx}
\usepackage{amsfonts}
\usepackage{url}
\usepackage{epstopdf}
\usepackage{color}
\usepackage{bm}
\usepackage[colorlinks=true,linkcolor=blue,citecolor=red]{hyperref}%
\usepackage{comment}

\def\qa{{q_{\rm a}}}
\def\qc{{q_{\rm c}}}
\def\qccrit{{q_{\rm c}^{\rm crit}}}
\def\Gammaa{\Gamma_{\rm a}}
\def\Gammac{\Gamma_{\rm c}}
\def\Gammar{\Gamma_{\rm r}}

\def\a{\mathrm{a}}
\def\c{\mathrm{c}}

\def\L{\mathcal L}

\def\pa{\partial\Omega}
\def\E{{\mathbb E}}
\def\P{{\mathbb P}}
\def\R{{\mathbb R}}
\def\T{{\mathcal T}}

\def\erfc{\mathrm{erfc}}
\def\erfcx{\mathrm{erfcx}}

\def\x{\bm{x}}

\begin{document}

\title{Surviving the Attack of the Clones}

\author{Denis~S.~Grebenkov}
 \email{denis.grebenkov@polytechnique.edu}
\affiliation{
Laboratoire de Physique de la Mati\`{e}re Condens\'{e}e, \\ 
CNRS -- Ecole Polytechnique, Institut Polytechnique de Paris, 91120 Palaiseau, France}

\date{\today}

\begin{abstract}
We consider a population dynamics model in which each diffusing
particle that hits a catalytic surface can split into two independent
copies (clones).  The particles of such a growing-in-size population
search in parallel for a hidden partially reactive target to trigger a
reaction event (e.g., a viral attack).  We investigate the statistics
of the fastest first-reaction time (FRT) among all the particles.  We
establish a nonlinear integral equation for the survival probability
and then analyze the associated probability density of the FRT and its
moments.  Lower and upper bounds on the mean FRT are then deduced in
terms of the system parameters (target reactivity, catalytic rate,
diffusivity, etc.).  Because autocatalytic replication can rapidly
increase the number of searchers, it can substantially accelerate the
diffusive search.  We solve the nonlinear equations numerically in a
basic geometric setting and reveal advantages and limitations on the
autocatalytic search.
\end{abstract}





\maketitle

\section{Introduction}
\label{sec:intro}

In many applications, a swarm of independently moving particles search
in parallel for a target to trigger a specific event
\cite{Alberts,Redner,Schuss,Metzler,Masoliver,Lindenberg,Grebenkov,Bressloff13,Bray13,Benichou14}.
Common biological examples are the injection of up to 300 million
sperm cells searching for an egg \cite{Reynaud15,Yang16}, or several
thousand neurotransmitters searching for a specific receptor to
transmit the signal between neighboring neurons
\cite{Berridge03,Reva21}.  The efficiency of this multi-particle
search is characterized by the extremal statistics of the fastest
first-passage time (fFPT) to the target.  In the particular case of
Brownian particles with a constant diffusivity $D$, the mean fFPT
scales with the number of particles $N$ as $L^2/(4D\ln N)$, where $L$
is the distance from the starting point to the target \cite{Weiss83}.
Higher-order moments, the limiting distribution, and the statistics of
the $k$-th fFPT were rigorously characterized
\cite{Schuss19,Lawley20a,Lawley20b,Lawley20c}, whereas limitations of
these results and their extensions to non-Brownian dynamics were
discussed \cite{Lawley20d,Lawley21,Grebenkov26f}.  In particular, the
respective roles of the starting points and entrance times onto the
mean fFPT and its distribution were inspected
\cite{Grebenkov20a,Madrid20,Grebenkov25a}.

In this paper, we look at the multi-particle search problem from a
different perspective.  Suppose that a single particle diffuses in a
medium that contains catalytic regions, on which the particle can be
replicated.  To what extent can such a cloning process accelerate the
diffusive search?  At first sight, one might expect that the optimal
strategy is for a single particle to travel directly to the target.
However, as the likelihood of following such direct trajectories is
very small, the mean FPT to the target is dominated by long
explorations of the medium.  A more advantageous strategy may thus be
to spend some time near the catalytic region, thereby increasing the
number of searchers through cloning before resuming the search by a
large swarm.

\begin{figure}
\begin{center}
\includegraphics[width=60mm]{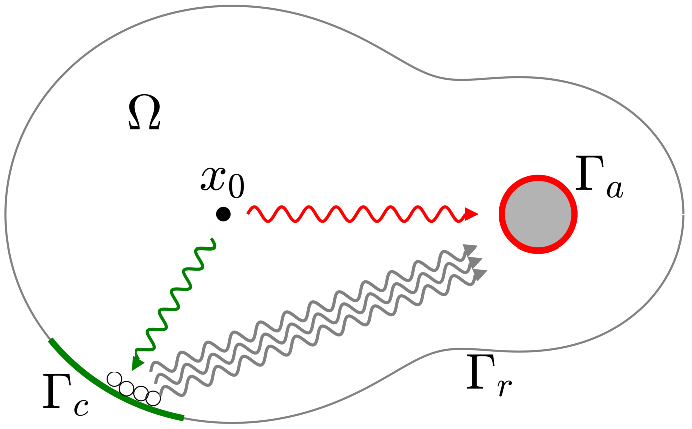} 
\end{center}
\caption{
Schematic view of a bounded domain $\Omega$, whose boundary $\pa$ is
split into a catalytic region $\Gammac$ (green), a target $\Gammaa$
(red), and the remaining reflecting part $\Gammar$ (gray).  A single
particle is released from a point $\x_0$ and diffuses in $\Omega$.
Some random trajectories reach the target directly (red wavy line),
whereas others first visit the catalytic region (green wavy line),
where the particle may split into two clones that can subsequently
split again.  These cloning events can rapidly increase the population
of the clones that search for the target in parallel (gray wavy
lines).  Note that both catalytic and target regions can be composed
of multiple pieces, which can be located either on the ``outer
frontier'', or ``inside the domain'' (in other words, the boundary
$\pa$ may be disconnected, as shown here).}
\label{fig:scheme}
\end{figure}

To address this optimization problem quantitatively, we consider a
broad class of diffusion-driven autocatalytic reactions on surfaces
\cite{Grebenkov26a,Grebenkov26b,Grebenkov26c}.  More specifically, let
$\Gammac \subset \pa$ denote the catalytic region on the boundary
$\pa$ of a bounded domain $\Omega \subset \R^d$
(Fig. \ref{fig:scheme}).  A single particle is released at time $0$
from a point $\x_0 \in \bar{\Omega} = \Omega \cup \pa$ and diffuses
with a constant diffusivity $D$.  When it hits $\Gammac$, it can be
transformed into two identical copies of itself (two clones).  This
replication event occurs with a small probability that is proportional
to the catalytic rate $\qc > 0$ (even though $\qc$ has units of
inverse length, it plays the role of a ``rate'' with respect to the
boundary local time, see \cite{Grebenkov26c,Grebenkov20} for more
details on the probabilistic description of this branching event).
After splitting, the original particle disappears, whereas two newborn
particles diffuse independently from the point of their birth, and
each of them can split upon hitting $\Gammac$ with the same rate
$\qc$.  In addition, we assign another subset $\Gammaa\subset \pa$ of
the boundary to be a target.  Any particle hitting the target can
react on it with a small probability that is proportional to the
reaction rate $\qa > 0$ (in the formal limit $\qa =
\infty$, the probability of reaction is $1$, meaning a perfectly
reactive target; note that $\qa D$ is usually called reactivity).  The
reaction event is understood here as the completion of the search
process.  We aim at characterizing the first-reaction time $\T$, i.e.,
the first-passage time to the reaction event on the target by any of
the existing particles.  Figuratively speaking, the target can be
considered as a ``base'', which is ``attacked'' by the diffusing
clones, and our goal is to compute the probability
\begin{equation}  \label{eq:S_def}
S(t|\x_0) = \P_{\x_0}\{ \T > t\} 
\end{equation}
of the base survival up to time $t$, and thus the distribution of the
random time $\T$ of the base ``destruction'' after the successful
attack.

The paper is organized as follows.  In Sec. \ref{sec:theory}, we
develop a general theory to describe the first-passage properties of
the stochastic dynamics of the above diffusion-reaction system.  In
particular, we derive a nonlinear integral equation for the survival
probability $S(t|\x_0)$ and investigate its long-time behavior.  We
also discuss the mean first-reaction time (MFRT) and its bounds in
terms of conventional MFRTs for a single particle.  In
Sec. \ref{sec:example}, we illustrate the first-passage properties for
a basic example of an interval $(0,L)$, with the catalytic region at
$0$ and the target at $L$.  While the diffusive dynamics of a single
particle is elementary and well-known, the nonlinear nature of
autocatalytic reactions still presents a major challenge in getting an
explicit solution even for the MFRT $T(\x_0) = \E_{\x_0}\{\T\}$.  We
therefore solve the nonlinear equation numerically to reveal the
impact of autocatalytic reactions onto the search process.  We also
present direct Monte Carlo simulations for validation purposes and
briefly discuss comparison with the conventional case of a fixed-size
population with $N$ independent searchers.  Section
\ref{sec:discussion} concludes the paper with the summary of main
results, further extensions and open problems.

\section{Theory}
\label{sec:theory}

Following \cite{Grebenkov26b,Grebenkov26c}, we aim at establishing an
integral equation for the survival probability $S(t|\x_0)$ by
inspecting the first splitting event.  For this purpose, we introduce
the single-particle propagator $P(\x,t|\x_0)$ that describes the
competition between the subsets $\Gammac$ and $\Gammaa$ for capturing
the particle and satisfies for any fixed starting point $\x_0 \in
\Omega$:
\begin{subequations}
\begin{align}  \label{eq:P_diff}
\partial_t P & = D \Delta P \quad (\x\in\Omega), \\
\partial_n P + \qc P & = 0 \quad (\x\in \Gammac), \\  \label{eq:P_qa}
\partial_n P + \qa P & = 0 \quad (\x\in \Gammaa), \\
\partial_n P & = 0 \quad (\x\in \Gammar), \\
P(\x,0|\x_0) & = \delta(\x-\x_0),
\end{align}
\end{subequations}
where $\Delta$ is the Laplace operator, $\partial_n$ is the normal
derivative oriented outwards the domain $\Omega$, $\Gammar = \pa
\backslash \overline{\Gammac \cup \Gammaa}$ is the reflecting
(passive) part of the boundary, and $\delta(\x-\x_0)$ is the Dirac
distribution that fixes the starting point $\x_0$ at time $0$.  The
symmetry of the propagator, $P(\x,t|\x_0) = P(\x_0,t|\x)$, allows one
to set the starting point $\x_0$ on the boundary $\pa$ as well.  When
the target is perfectly reactive (i.e., it is found instantly upon the
first arrival), one can formally set $\qa = \infty$ that transforms
the Robin boundary condition (\ref{eq:P_qa}) into the Dirichlet
condition $P = 0$ on $\Gammaa$.
The propagator $P(\x,t|\x_0)$ admits the spectral expansion
\cite{Gardiner,Risken},
\begin{equation}  \label{eq:P_spectral}
P(\x,t|\x_0) = \sum\limits_{k=0}^\infty u_k(\x) \, u_k(\x_0) \, e^{-Dt\lambda_k},
\end{equation}
where $\lambda_k$ and $u_k(\x)$ are the Laplacian eigenvalues and
eigenfunctions satisfying:
\begin{subequations}
\begin{align}
-\Delta u_k & = \lambda_k u_k \quad (\x\in \Omega), \\  \label{eq:uk_qc}
\partial_n u_k + \qc u_k & = 0 \quad (\x\in\Gammac), \\
\partial_n u_k + \qa u_k & = 0 \quad (\x\in\Gammaa), \\
\partial_n u_k & = 0 \quad (\x\in\Gammar).
\end{align}
\end{subequations}
The eigenvalues are enumerated by a nonnegative index $k$ in an
increasing order, $0 < \lambda_0 \leq \lambda_1 \leq \lambda_2 \leq
\cdots \nearrow +\infty$, whereas the eigenfunctions form a complete
orthonormal basis of the functional space $L^2(\Omega)$ of
square-integrable functions \cite{Levitin}:
\begin{equation}
\int\limits_{\Omega} d\x \, u_j(\x) \, u_k(\x) = \delta_{j,k} .
\end{equation}

Let us denote by $\tau_{\a}$ the first-reaction time of the first
particle on $\Gammaa$ and by $\tau_{\c}$ its first-splitting time on
$\Gammac$.  For a given time $t$, one of the three mutually exclusive
events occurs: (i) $t < \min\{\tau_{\c},\tau_{\a}\}$ (no splitting, no
reaction up to time $t$), (ii) $\tau_{\c} < \min\{t,\tau_{\a}\}$
(splitting of the first particle before $t$ and before its reaction on
$\Gammaa$), and (iii) $\tau_{\a} < \min\{t,\tau_{\c}\}$ (reaction on
$\Gammaa$ before splitting and before $t$).  The probability of no
reaction up to time $t$ for {\it any} particle in the population can
thus be written as a renewal equation:
\begin{align}  \label{eq:S_integral}
S(t|\x_0) & = S_\a(t|\x_0) \\  \nonumber
& + \int\limits_{\Gammac} d\x \int\limits_0^t dt' \, \qc D P(\x,t'|\x_0) \, S^2(t-t'|\x).
\end{align}
In fact, after the first splitting event, the process restarts with
two independent descendants, which explains the nonlinear term
$S^2(t-t'|\x)$.  Here
\begin{align}  
S_{\a}(t|\x_0) & = \P_{\x_0}\{ t < \min\{\tau_{\c} ,\tau_{\a}\}\} 
 =  \int\limits_{\Omega} d\x \, P(\x,t'|\x_0)
\end{align}
is the probability of having neither catalytic splitting, nor reaction
up to time $t$ (option 1).  In turn, the second option is represented
via the second term of Eq. (\ref{eq:S_integral}), where $-D\partial_n
P(\x,t'|\x_0)|_{\Gammac} = \qc D P(\x,t'|\x_0)$ is the probability
flux density that determines the joint probability density of the
splitting time and the location of the splitting event on $\Gammac$.
As two newborn particles start from this location $\x$ and diffuse
independently, the probability of no reaction up to the remaining time
$t-t'$ is $S^2(t-t'|\x)$.  The second term in
Eq. (\ref{eq:S_integral}) averages over all possible splitting times
$t'\in (0,t)$ and all possible splitting locations $\x\in\Gammac$.
Importantly, the factor $S^2(t-t'|\x)$ in Eq. (\ref{eq:S_integral}) is
not a mean-field closure; it follows exactly from the branching
property.  The integral equation (\ref{eq:S_integral}) is the first
main result of this paper.

In analogy to derivations presented in \cite{Grebenkov26b}, one can
easily check that the solution of the integral equation
(\ref{eq:S_integral}) satisfies an equivalent partial differential
equation (PDE) problem:
\begin{subequations}  \label{eq:S_PDE}
\begin{align}  \label{eq:S_diff}
\partial_t S & = D \Delta S \quad (\x_0\in\Omega), \\  \label{eq:S_qc}
\partial_n S + \qc S & = \qc S^2  \quad (\x_0\in \Gammac), \\  \label{eq:S_qa}
\partial_n S + \qa S & = 0 \quad (\x_0\in \Gammaa), \\
\partial_n S & = 0 \quad (\x_0\in \Gammar), \\
S(0|\x_0) & = 1.
\end{align}
\end{subequations}
In fact, as splitting events occur exclusively on the catalytic region
$\Gammac$ of the boundary, the survival probability $S(t|\x_0)$
satisfies the standard (backward) diffusion equation
(\ref{eq:S_diff}), which is however subject to the {\it nonlinear}
Robin-type boundary condition (\ref{eq:S_qc}).  Note that a similar
PDE was established in \cite{Grebenkov26b} for the generating function
of the population size; however, these equations are different that is
reflected in the distinct properties of the survival probability and
the generating function, as discussed below.

Once the survival probability is found, the probability density of the
FRT $\T$,
\begin{equation}
J(t|\x_0) = - \partial_t S(t|\x_0), 
\end{equation}
can be obtained via a direct numerical calculation of the time
derivative.  An alternative computation consists in solving a linear
integral equation, which is obtained by evaluating the time derivative
of Eq. (\ref{eq:S_integral}):
\begin{align*}  
J(t|\x_0) & = -\partial_t S_\a(t|\x_0) - \qc D \int\limits_{\Gammac} d\x \biggl\{ P(\x,t|\x_0) \underbrace{S^2(0|\x)}_{=1} \\
& - 2\int\limits_0^t dt' \, P(\x,t'|\x_0) \, S(t-t'|\x) \, J(t-t'|\x)\biggr\}.
\end{align*}
To proceed, we use the following identity
\begin{align*}
\partial_t S_\a(t|\x_0) & = -\int\limits_{\Gammac} d\x \, \qc D \, P(\x,t|\x_0) \\
& + \int\limits_{\Gammaa} d\x \, D\partial_n P(\x,t|\x_0),
\end{align*}
which is deduced by evaluating the integral of Eq. (\ref{eq:P_diff})
over $\x\in\Omega$ and applying the Green's formula.  As a
consequence, defining
\begin{align} \nonumber
J_\a(t|\x_0) & = - \partial_t S_\a(t|\x_0) - \qc D \int\limits_{\Gammac} d\x\, P(\x,t|\x_0) \\
& = \int\limits_{\Gammaa} d\x \, (-D\partial_n P(\x,t|\x_0)), 
\end{align}
we have
\begin{align}\nonumber  
J(t|\x_0) & = J_\a(t|\x_0) + 2\qc D \int\limits_{\Gammac} d\x 
\int\limits_0^t dt' \, P(\x,t'|\x_0) \\
& \times  S(t-t'|\x) \, J(t-t'|\x).
\end{align}
This is a linear integral equation that determines $J(t|\x_0)$, once
the survival probability $S(t|\x_0)$ is known.

From the probability density, one can determine all the moments of the
first-reaction time $\T$.  In particular, the mean value,
\begin{equation}  \label{eq:Tmean_def}
T(\x_0) = \E_{\x_0}\{ \T\} = \int\limits_0^\infty dt \, t \, J(t|\x_0) = \int\limits_0^\infty dt \, S(t|\x_0) ,
\end{equation}
is obtained by integrating Eq. (\ref{eq:S_integral}) over $t$ from $0$
to infinity:
\begin{equation}  \label{eq:Tmean_integral}
T(\x_0) = T_\a(\x_0) + \qc D \int\limits_{\Gammac} d\x \, \tilde{P}(\x,0|\x_0) T^{(2)}(\x),
\end{equation}
where 
\begin{equation}
T_\a(\x_0) = \int\limits_0^\infty dt \, S_{\a}(t|\x_0)
\end{equation}
is the MFRT on either of regions $\Gammaa$ and $\Gammac$, 
\begin{equation}  \label{eq:T2_def}
T^{(2)}(\x) = \int\limits_0^\infty dt \, S^2(t|\x),
\end{equation}
and we used the Laplace transform of the propagator,
\begin{equation}
\tilde{P}(\x,p|\x_0) = \int\limits_0^\infty dt \, e^{-pt} \, P(\x,t|\x_0),
\end{equation}
denoted by tilde and evaluated at $p = 0$.
Alternatively, one can integrate Eqs. (\ref{eq:S_PDE})
over $t$ from $0$ to infinity to get
\begin{subequations}  \label{eq:MFPT_PDE}
\begin{align}  \label{eq:MFPT_eq}
D \Delta T &= - 1 \quad (\x_0\in\Omega), \\  \label{eq:MFPT_qc}
\partial_n T + \qc T & = \qc T^{(2)}  \quad (\x_0\in \Gammac), \\  \label{eq:MFPT_qa}
\partial_n T + \qa T & = 0 \quad (\x_0\in \Gammaa), \\
\partial_n T & = 0 \quad (\x_0\in \Gammar).
\end{align}
\end{subequations}
In sharp contrast to the conventional MFRT, the present problem is not
closed at the level of the mean because the boundary condition
involves the unknown function $T^{(2)}(\x)$.  In other words, one
still needs to compute the survival probability $S(t|\x_0)$ to be able
to evaluate $T(\x_0)$.  This is the consequence of the autocatalytic
dynamics and eventual correlations that emerge due to a simultaneous
birth of two particles via cloning.

\subsection{Long-time behavior}

According to Eq. (\ref{eq:P_spectral}), both $P(\x,t|\x_0)$ and
$S_{\a}(t|\x_0)$ exhibit the exponential decay at long times that is
controlled by the principal Laplacian eigenvalue $\lambda_0$.
Consequently, the survival probability $S(t|\x_0)$ is expected to
exhibit the same exponential decay.  This expectation can be confirmed
by taking the Laplace transform of Eq. (\ref{eq:S_integral}):
\begin{equation}
\tilde{S}(p|\x_0) = \tilde{S}_{\a}(p|\x_0) + \qc D \int\limits_{\Gammac} d\x \, \tilde{P}(\x,p|\x_0) \,\L\{ S^2(t|\x)\}(p),
\end{equation}
where both $\L$ and tilde denote the Laplace transform.  Both
$\tilde{S}_{\a}(p|\x_0)$ and $\tilde{P}(\x,p|\x_0)$ as functions of
$p$ have the largest pole at $-D\lambda_0$.  Assuming the long-time
behavior $S(t|\x_0) \propto e^{-\nu t}$ with some rate $\nu > 0$, the
largest pole of $\L\{ S^2(t|\x)\}(p)$ is then $-2\nu$.  Matching the
dominant poles on both sides of the equation yields $\nu= D\lambda_0$.
Moreover, since $\L\{ S^2(t|\x)\}(p)$ has no pole at $-\nu$, we
conclude that
\begin{equation}  \label{eq:St_long}
S(t|\x_0) \simeq A \, u_0(\x_0) \, e^{-Dt\lambda_0}  \qquad (t\to \infty),
\end{equation}
with
\begin{equation}
A = \int\limits_{\Omega} d\x \, u_0(\x)  +  \qc D \int\limits_{\Gammac} d\x \, u_0(\x) \int\limits_0^\infty dt \, e^{Dt\lambda_0} \,S^2(t|\x),
\end{equation}
where the last factor in the second term is the evaluation of the
residue at $p = -\nu = -D\lambda_0$.  While the asymptotic form of
$S(t|\x_0)$ is fully determined by the principal eigenmode of the
Laplace operator, the prefactor $A$ depends on $S^2(t|\x)$, integrated
over the entire range of times.

Interestingly, at long times the catalytic region $\Gammac$ affects
the decay rate exactly as if it were an additional partially reactive
target (with reaction rate $\qc$).  According to the standard
variational principle \cite{Levitin}, one has
\begin{equation}  \label{eq:lambda0_var}
\lambda_0 = \inf\limits_{v\in L^2(\Omega) \atop \|v\|_{L^2(\Omega)} = 1} \left\{ \int\limits_\Omega |\nabla v|^2
+ \qc \int\limits_{\Gammac} |v|^2 + \qa \int\limits_{\Gammaa} |v|^2 \right\}.
\end{equation}
This relation immediately implies that $\lambda_0$ monotonously
increases with both $\qa$ and $\qc$.  Equation (\ref{eq:lambda0_var})
provides a simple variational interpretation of the search
acceleration: increasing either the target reactivity or the catalytic
rate raises the principal eigenvalue, thereby shortening the
characteristic decay time of the survival probability.

\subsection{Dual representation}

As discussed in \cite{Grebenkov26c} in the context of the population
size dynamics, an implementation of the boundary condition through the
integral term in Eq.  (\ref{eq:S_integral}) allows one to derive
alternative but equivalent integral representations.  For this
purpose, let us consider the single-particle propagator
$P_0(\x,t|\x_0)$ with Neumann boundary condition on $\Gammac$:
\begin{subequations}
\begin{align}  \label{eq:P0_diff}
\partial_t P_0 & = D \Delta P_0 \quad (\x\in\Omega), \\
\partial_n P_0  & = 0 \quad (\x\in \Gammac), \\  \label{eq:P0_qa}
\partial_n P_0 + \qa P_0 & = 0 \quad (\x\in \Gammaa), \\
\partial_n P_0 & = 0 \quad (\x\in \Gammar), \\
P_0(\x,0|\x_0) & = \delta(\x-\x_0).
\end{align}
\end{subequations}
In other words, this propagator describes diffusion in the presence of
a partially reactive target but without any autocatalytic dynamics (as
if $\qc = 0$).
Following the same argument as in \cite{Grebenkov26c}, we transform
Eq. (\ref{eq:S_integral}) into
\begin{align}  \nonumber
S(t|\x_0) & = S_0(t|\x_0) - \qc D \int\limits_{\Gammac} d\x \int\limits_0^t dt' \,  P_0(\x,t'|\x_0) \\  \label{eq:S_integral2}
& \times \bigl(S(t-t'|\x) - S^2(t-t'|\x)\bigr),
\end{align}
where
\begin{equation}
S_0(t|\x_0) = \int\limits_{\Omega} d\x \, P_0(\x,t|\x_0).
\end{equation}
This alternative representation presents the second main result of the
paper and offers two practical advantages.  First, the propagator
$P_0(\x,t|\x_0)$ and the survival probability $S_0(t|\x_0)$ treat the
catalytic region as purely reflecting, i.e., they do not depend on the
catalytic rate $\qc$.  Second, this integral equation turns out to be
more suitable for a numerical solution (see Appendix
\ref{sec:numerics}).

From Eq. (\ref{eq:S_integral2}), we also deduce the convolution-type
equation
\begin{align}  \nonumber
J(t|\x_0) & = J_0(t|\x_0) + \qc D \int\limits_{\Gammac} d\x \int\limits_0^t dt' \,  P_0(\x,t'|\x_0) \\  \label{eq:J_integral2}
& \times \bigl(2S(t-t'|\x) - 1\bigr) J(t-t'|\x),
\end{align}
as well as
\begin{align}  \nonumber
T(\x_0) & = T_0(\x_0) - \qc D \int\limits_{\Gammac} d\x \, \tilde{P}_0(\x,0|\x_0) \\  \label{eq:T_integral2}
& \times \int\limits_0^\infty dt \bigl(S(t|\x) - S^2(t|\x)\bigr),
\end{align}
with
\begin{equation}
J_0(t|\x_0) = -\partial_t S_0(t|\x_0), \quad T_0(\x_0) = \int\limits_0^\infty dt \, S_0(t|\x_0).
\end{equation}

An immediate consequence of the integral equations
(\ref{eq:S_integral}, \ref{eq:S_integral2}) is the two-sided bound on
the survival probability:
\begin{equation}
S_{\a}(t|\x_0) \leq S(t|\x_0) \leq S_0(t|\x_0).
\end{equation}
The same bound holds for the MFRT (and higher-order moments):
\begin{equation}  \label{eq:T_bounds}
T_{\a}(\x_0) \leq T(\x_0) \leq T_0(\x_0).
\end{equation}
The upper bound has a simple physical interpretation: autocatalytic
replication can only create additional searchers and therefore cannot
make the search less efficient.  In fact, if a particle has reached
the catalytic region, an eventual splitting event may increase the
number of particles and thus speed up the search, as compared to the
case $\qc = 0$ when the same particle is just reflected from $\Gammac$
and continues searching alone.  In turn, the lower bound means that if
the catalytic region $\Gammac$ was partially reactive (with the rate
$\qc$), then a reaction event on $\Gammac$ would mean the completion
of the search process; however, when $\Gammac$ is catalytic, such a
reaction event on $\Gammac$ leads to splitting of the particle, and
the search process continues.

The lower bound may appear counterintuitive, particularly for very
large catalytic rates: despite the fact that the population of
particles grows extremely fast (see \cite{Grebenkov26a} and
Sec. \ref{sec:qccrit}), the MFRT $T(\x_0)$ remains limited by
$T_\a(\x_0)$ and cannot thus significantly decrease, especially if the
starting point $\x_0$ is not located on the catalytic region.  In
fact, if $\x_0 \notin \Gammac$, the lower bound $T_{\a}(\x_0)$
approaches a finite limit as $\qc \to \infty$.  The rationale for this
limited improvement is elementary: in order to initiate such a rapid
growth, the first particle needs to reach the catalytic region, and
the required mean time is related to $T_{\a}(\x_0)$.  In other words,
the initial journey toward the catalytic region constitutes an
irreducible bottleneck that limits the achievable speed-up.
We will illustrate this behavior by numerical examples in
Sec. \ref{sec:example}.

\subsection{Regimes of low and high catalytic rates}
\label{sec:qc_regimes}

To determine the dependence of the survival probability and the MFRT
on the catalytic rate $\qc$, one needs to solve the nonlinear integral
equation (\ref{eq:S_integral}) or the PDE problem (\ref{eq:S_PDE}).
While its numerical study for a simple geometric setting is postponed
to Sec. \ref{sec:example}, we briefly sketch some basic properties and
highlight the eventual difficulties.

In the limit of small $\qc$, the boundary condition (\ref{eq:S_qc}) is
close to the Neumann condition, and the survival probability
$S(t|\x_0)$ should be close to $S_0(t|\x_0)$.  In fact, as cloning
events are rare, the catalytic region mainly reflects the particles.
More formally, Eq. (\ref{eq:S_integral2}) implies immediately that
$S_0(t|\x_0)$ is the leading-order approximation for $S(t|\x_0)$ as
$\qc \to 0$.  Moreover, one can apply a perturbation approach to get
higher-order corrections in terms of $S_0(t|\x_0)$ and
$P_0(\x,t|\x_0)$.  Similarly, Eq. (\ref{eq:T_integral2}) implies a
two-term approximation for the MFRT:
\begin{align} \nonumber
T(\x_0) & = T_0(\x_0) - \qc D \int\limits_{\Gammac} d\x \, P_0(\x,t|\x_0) \\  \label{eq:T_qcsmall}
& \times \int\limits_0^\infty dt \bigl(S_0(t|\x) - S_0^2(t|\x)\bigr) + O(q_{\c}^2).
\end{align}

The opposite limit of large $\qc$ is considerably more subtle; in
fact, even for the conventional survival probability $S_{\a}(t|\x_0)$,
the limit $\qc\to\infty$ of the Robin boundary condition is known to
be singular.  Dividing the boundary condition (\ref{eq:S_qc}) by $\qc$
and taking $\qc \to \infty$ lead to the limiting boundary condition
$S(t|\x_0) = S^2(t|\x_0)$ on $\Gammac$.  Since $S(t|\x_0)$ is bounded
above by $S_0(t|\x_0)$, which is strictly less than $1$ for any $t >
0$, one can discard the solution $S(t|\x_0) = 1$ on $\Gammac$ and thus
get the Dirichlet boundary condition $S(t|\x_0) = 0$ on $\Gammac$ in
the limit.  In other words, $S(t|\x_0)$ is expected to become closer
and closer to $S_{\a}(t|\x_0)$ as $\qc \to \infty$.  Similarly, the
MFRT $T(\x_0)$ is expected to be close to $T_{\a}(\x_0)$, which
satisfies the conventional Poisson equation:
\begin{subequations}  \label{eq:Ta_PDE}
\begin{align}  
D \Delta T_\a &= - 1 \quad (\x_0\in\Omega), \\ 
\partial_n T_\a + \qc T_\a & = 0  \quad (\x_0\in \Gammac), \\ 
\partial_n T_\a + \qa T_\a & = 0 \quad (\x_0\in \Gammaa), \\
\partial_n T_\a & = 0 \quad (\x_0\in \Gammar).
\end{align}
\end{subequations}
The latter admits a standard perturbation analysis.  In fact,
substituting $T_\a(\x_0) = v_0(\x_0) + v_1(\x_0)/\qc + \cdots$ into
this equation, one sees that $v_0$ is the MFRT with the Dirichlet
boundary condition $v_0 = 0$ on $\Gammac$, whereas the first-order
correction $v_1$ satisfies
\begin{subequations}  \label{eq:v1_PDE}
\begin{align}  
\Delta v_1 &= 0 \quad (\x_0\in\Omega), \\ 
v_1 & = -\partial_n v_0 \quad (\x_0\in \Gammac), \\ 
\partial_n v_1 + \qa v_1 & = 0 \quad (\x_0\in \Gammaa), \\
\partial_n v_1 & = 0 \quad (\x_0\in \Gammar).
\end{align}
\end{subequations}
In particular, this perturbative analysis implies that if the starting
point is located at $\Gammac$, one has $T_{\a}|_{\Gammac} \approx
q_{\c}^{-1} (-\partial_n v_0)|_{\Gammac}$ in the leading order.  A
similar asymptotic behavior is expected for the MFRT $T(\x_0)$.  We
emphasize that the above heuristic arguments remain formal and require
rigorous mathematical justification, which lies beyond the scope of
the present work.

\subsection{Moderate efficiency of autocatalytic reactions}
\label{sec:qccrit}

The competition between cloning events on $\Gammac$ and reaction
events on $\Gammaa$ and the consequent long-time behavior of the mean
population size $N(t|\x_0)$ was investigated in \cite{Grebenkov26a}.
In particular, it was shown the existence of a critical catalytic rate
$\qccrit$ that distinguished three scenarios: (i) the subcritical
regime $\qc < \qccrit$, in which $N(t|\x_0)$ vanishes at long times;
(ii) the critical regime $\qc = \qccrit$ when $N(t|\x_0)$ reaches a
nonzero steady-state; and (iii) the supercritical regime $\qc >
\qccrit$ when $N(t|\x_0)$ grows exponentially fast as $t\to\infty$.  A
mathematical framework to determine the critical catalytic rate
$\qccrit$ and its dependence on $\qa$ and on the geometric
configuration was established.

At first glance, it may be surprising that the critical catalytic rate
has not yet appeared in our former analysis.  For instance, the
survival probability $S(t|\x_0)$ exhibits the long-time exponential
decay (\ref{eq:St_long}) with the rate $\lambda_0$, which smoothly
depends on the catalytic rate $\qc$ (as it follows from the
variational principle (\ref{eq:lambda0_var})).  In other words, the
properties of the FRT do not almost change when $\qc$ crosses the
critical value $\qccrit$.  This apparent paradox has two complementary
explanations.  (i) In mathematical terms, the long-time behavior of
the mean population size is controlled by another eigenvalue,
$\lambda_0^-$, that corresponds to the Laplacian eigenvalue problem,
in which $+\qc$ in the Robin boundary condition (\ref{eq:uk_qc}) is
replaced by the {\it negative} parameter $-\qc$.  As discussed in
\cite{Grebenkov26a}, the long-time behavior $N(t|\x_0)
\propto e^{-Dt\lambda_0^-}$ is very sensitive to the sign of
$\lambda_0^-$, which vanishes at $\qc = \qccrit$ and thus
distinguishes the above three regimes.  In contrast, the long-time
behavior of the survival probability $S(t|\x_0)$ is controlled by the
eigenvalue $\lambda_0$, which smoothly depends on $\qc$.  (ii) In the
analysis of the population dynamics, the stochastic process was not
stopped upon the first-reaction time, and the long-time behavior of
$N(t|\x_0)$ results from the competition between (usually many)
cloning and reaction events.  In turn, the analysis of the FRT imposes
to stop the stochastic process after the first reaction event so that
the effect of the catalytic rate $\qc$ on the FRT is substantially
weaker.

\section{Example of an interval}
\label{sec:example}

To illustrate the effect of autocatalytic dynamics onto the
first-reaction time distribution, we consider a basic setting of an
interval, $\Omega = (0,L)$, with the catalytic endpoint $\Gammac =
\{0\}$ and the target $\Gammaa = \{L\}$.  Although we focus on the
interval for clarity, the analysis extends straightforwardly to
concentric annuli and spherical shells.  In fact, the rotational
invariance of these domains makes all computations effectively
one-dimensional so that only technical details would change (see,
e.g., \cite{Grebenkov26c} for a concentric annulus).  In this paper,
we focus on the interval to keep the presentation as transparent as
possible.

In this case, the critical catalytic rate is \cite{Grebenkov26a} 
\begin{equation}
\qccrit = \frac{\qa}{1 + \qa L}  \,.  
\end{equation}
As discussed in Sec. \ref{sec:qccrit}, this threshold distinguishes
three asymptotic regimes for the mean population size but it is
expected to play no role in the analysis of the FRT distribution.  We
will confirm this statement by inspecting different values of the
catalytic rate $\qc$.
In the following numerical illustrations, we set $L = 1$ and $D = 1$
to fix length and time units.

\subsection{Survival probability and PDF}
\label{sec:int_survival}

Since the catalytic boundary $\Gammac$ of an interval is reduced to a
single point, the integral equations (\ref{eq:S_integral},
\ref{eq:S_integral2}) are simplified as
\begin{equation}  \label{eq:S_interval}
S(t|x_0) = S_\a(t|x_0) + \qc D \int\limits_0^t dt' \, P(0,t'|x_0)\, S^2(t-t'|0)
\end{equation}
and
\begin{align} \nonumber
S(t|x_0) & = S_0(t|x_0) - \qc D \int\limits_0^t dt' \, P_0(0,t'|x_0) \\   \label{eq:Sinterval_eq}
& \times \bigl(S(t-t'|0) - S^2(t-t'|0)\bigr).
\end{align}
Either convolution equation can be solved numerically for $x_0 = 0$ to
find $S(t|0)$, from which the survival probability for any $x_0 > 0$
follows.  Results were independently verified using two different
numerical solvers (see Appendix \ref{sec:numerics}).  Details of both
numerical methods for solving these equations, as well as the explicit
formulas for $S_\a(t|x_0)$, $P(0,t|x_0)$, $S_0(t|x_0)$ and
$P_0(0,t|x_0)$, are presented in Appendix \ref{sec:numerics} (note
that the Laplace transform does not help much in our situation due to
the nonlinear factor like $S^2(t-t'|0)$).
For another independent verification, we also present the results of
Monte Carlo simulations (see Appendix \ref{sec:numerics}).  In order
to reveal the role of autocatalytic dynamics, we will generally
compare $S(t|0)$ and the related quantities $J(t|0)$ and $T(0)$ to
their counterparts $S_0(t|0)$, $J_0(t|0)$ and $T_0(0)$ corresponding
to $\qc = 0$, i.e., without autocatalytic dynamics.

\begin{figure}
\begin{center}
\includegraphics[width=88mm]{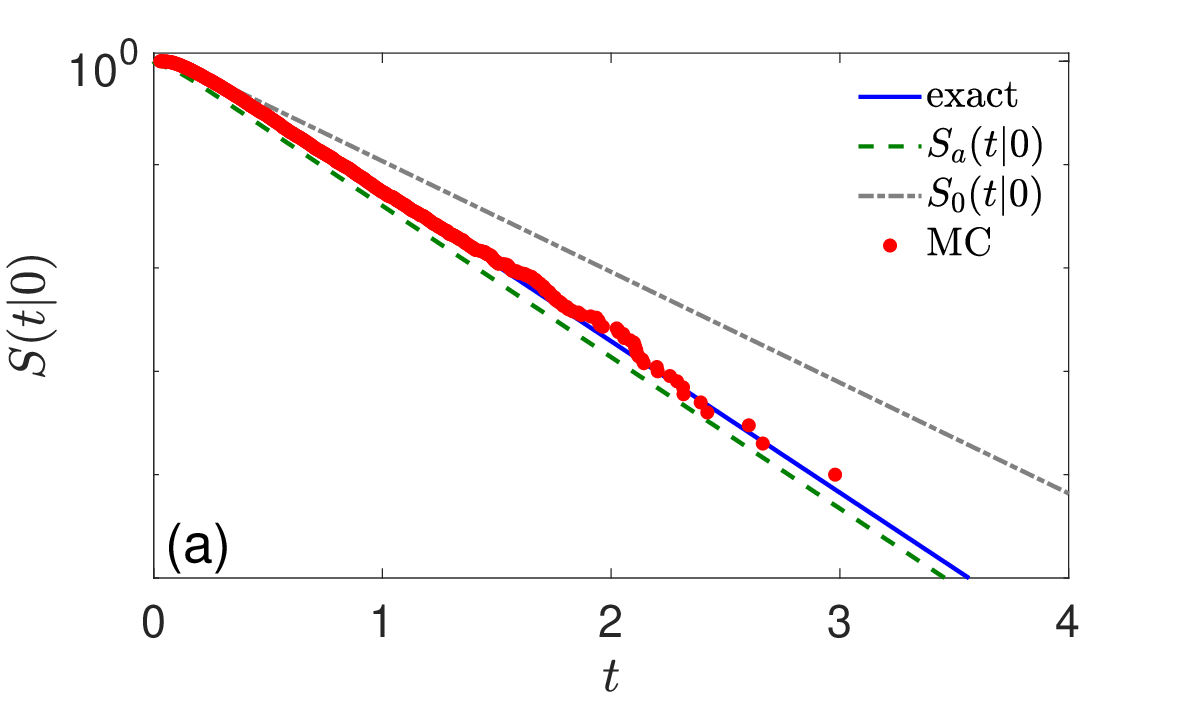} 
\includegraphics[width=88mm]{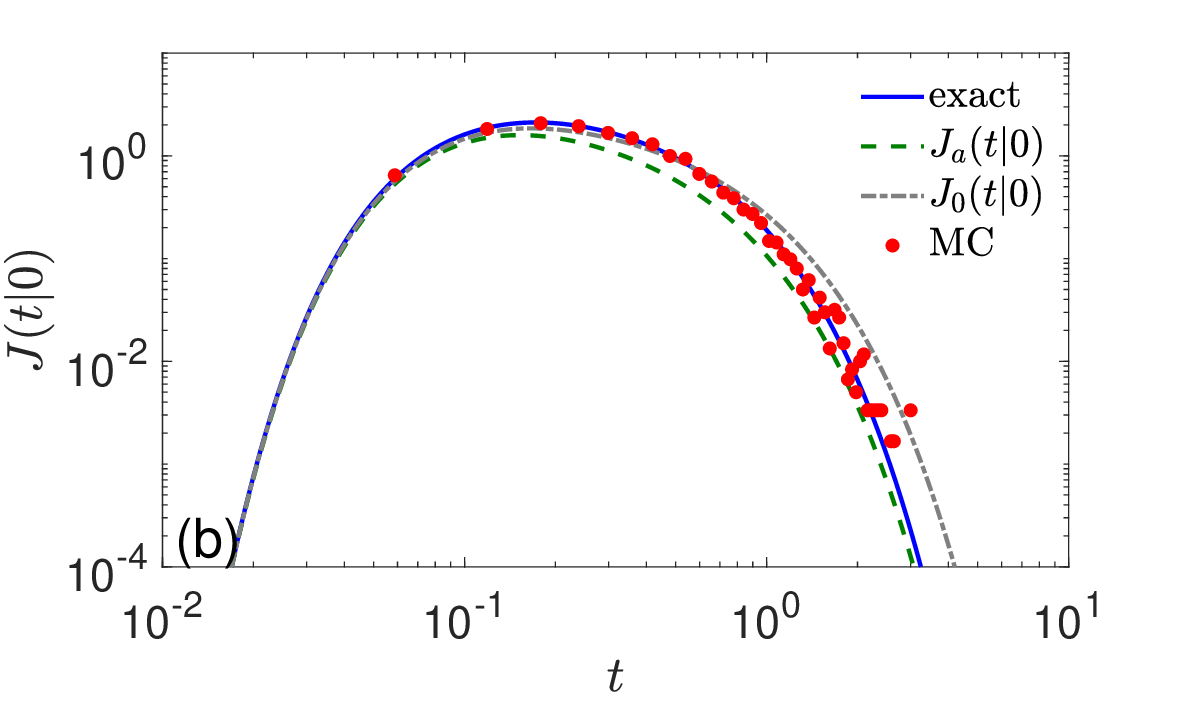} 
\end{center}
\caption{
Survival probability $S(t|0)$ [panel (a)] and the probability density
$J(t|0)$ of the FRT [panel (b)] for the unit interval ($L = 1$), with
$D = 1$, $\qa = \infty$, and $\qc = 0.5$.  Solid line presents the
exact quantities obtained via a numerical solution of the integral
equation (\ref{eq:S_integral2}).  Dashed and dash-dotted lines present
$S_{\a}(t|0)$ and $S_0(t|0)$ given by truncated spectral expansions.
Filled circles show the results of Monte Carlo simulations with $10^4$
realizations and the discretization step $a = 0.005$. }
\label{fig:SJ_q05}
\end{figure}

We begin by presenting the survival probability $S(t|0)$ and the
associated probability density $J(t|0)$ of the FRT on a perfect target
($\qa = \infty$), for which $\qccrit = 1$.  Figure \ref{fig:SJ_q05}
shows these quantities for $\qc = 0.5$ that corresponds to the
subcritical regime.  For comparison, the lower and upper bounds
$S_\a(t|0)$ and $S_0(t|0)$ are also drawn by dashed and dash-dotted
lines.  One sees that the lower bound $S_\a(t|0)$ is much closer to
the actual survival probability at moderate and long times.  This is
not surprising given that both functions decay at long times with the
same decay rate $D\lambda_0$ (in turn, the survival probability
$S_0(t|0)$ decays with a slower rate that corresponds to the passive
endpoint $0$).  The separation between the two bounds becomes even
more pronounced at larger $\qc = 5$ (Fig. \ref{fig:SJ_q5}).  In turn,
as shown in Appendix \ref{sec:Ashort} that the short-time asymptotic
behavior of $S(t|0)$ is close to $S_0(t|0)$ (not presented).  This
behavior is clearly visible on the panel (b) for the probability
density $J(t|0)$, which is close to $J_0(t|0)$ at short times.

\begin{figure}
\begin{center}
\includegraphics[width=88mm]{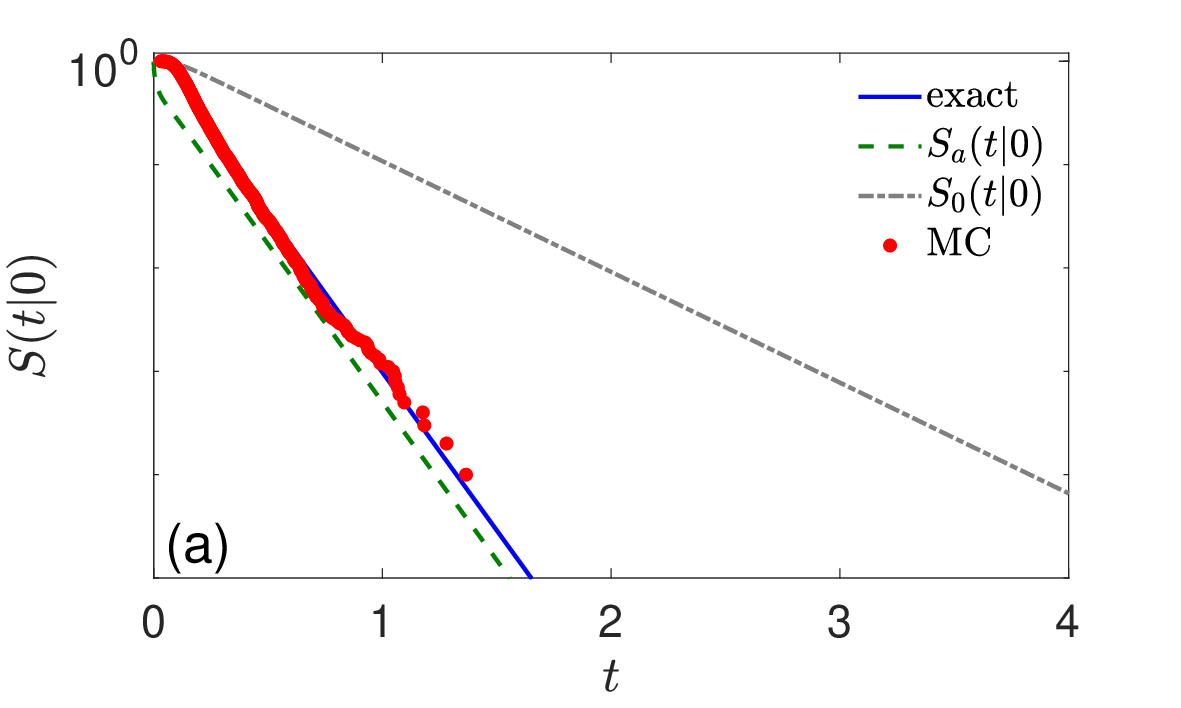} 
\includegraphics[width=88mm]{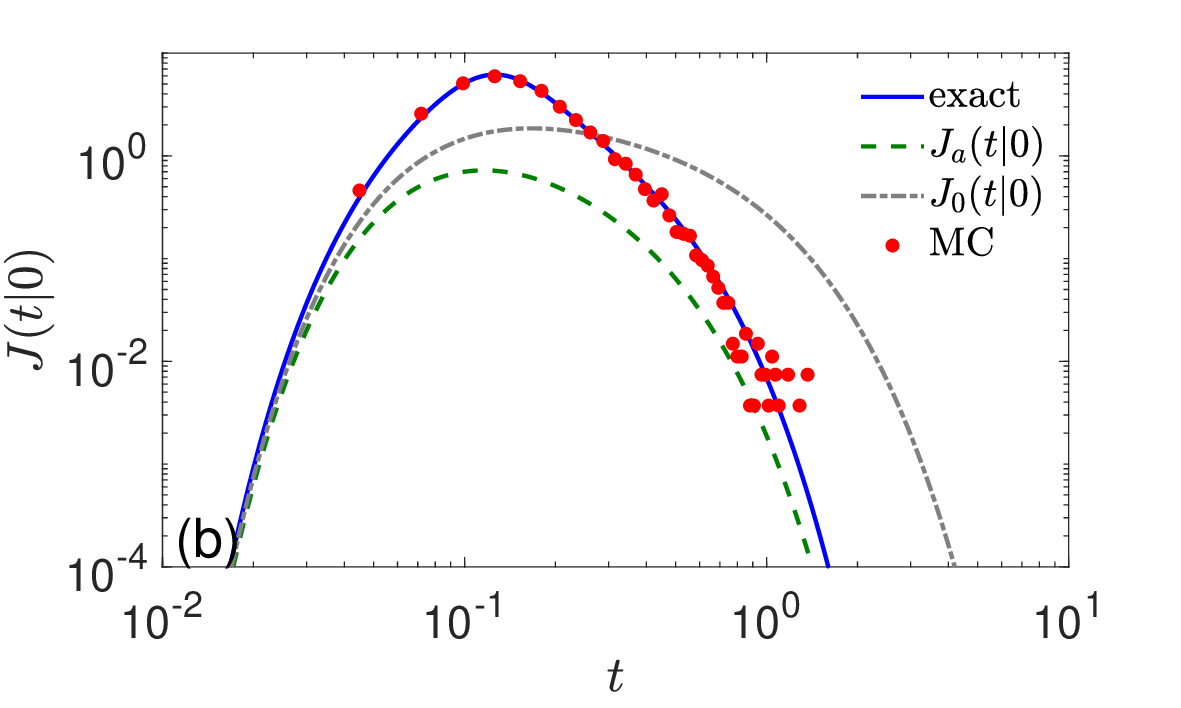} 
\end{center}
\caption{
Survival probability $S(t|0)$ [panel (a)] and the probability density
$J(t|0)$ of the FPT [panel (b)] for the unit interval ($L = 1$), with
$D = 1$, $\qa = \infty$, $\qc = 5$.  Solid line presents the exact
quantities obtained via a numerical solution of the integral equation
(\ref{eq:S_integral2}).  Dashed and dash-dotted lines present
$S_{\a}(t|0)$ and $S_0(t|0)$ given by truncated spectral expansions.
Filled circles show the results of Monte Carlo simulations with $10^4$
realizations and the discretization step $a = 0.005$. }
\label{fig:SJ_q5}
\end{figure}

\begin{figure}
\begin{center}
\includegraphics[width=88mm]{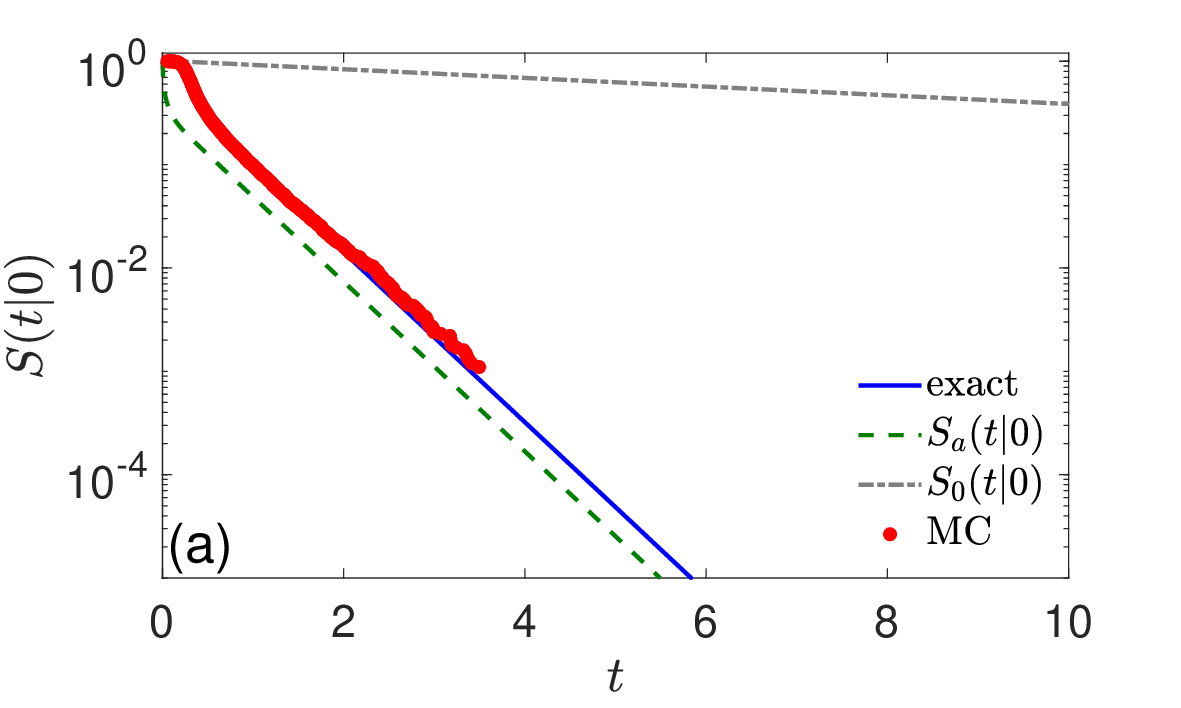} 
\includegraphics[width=88mm]{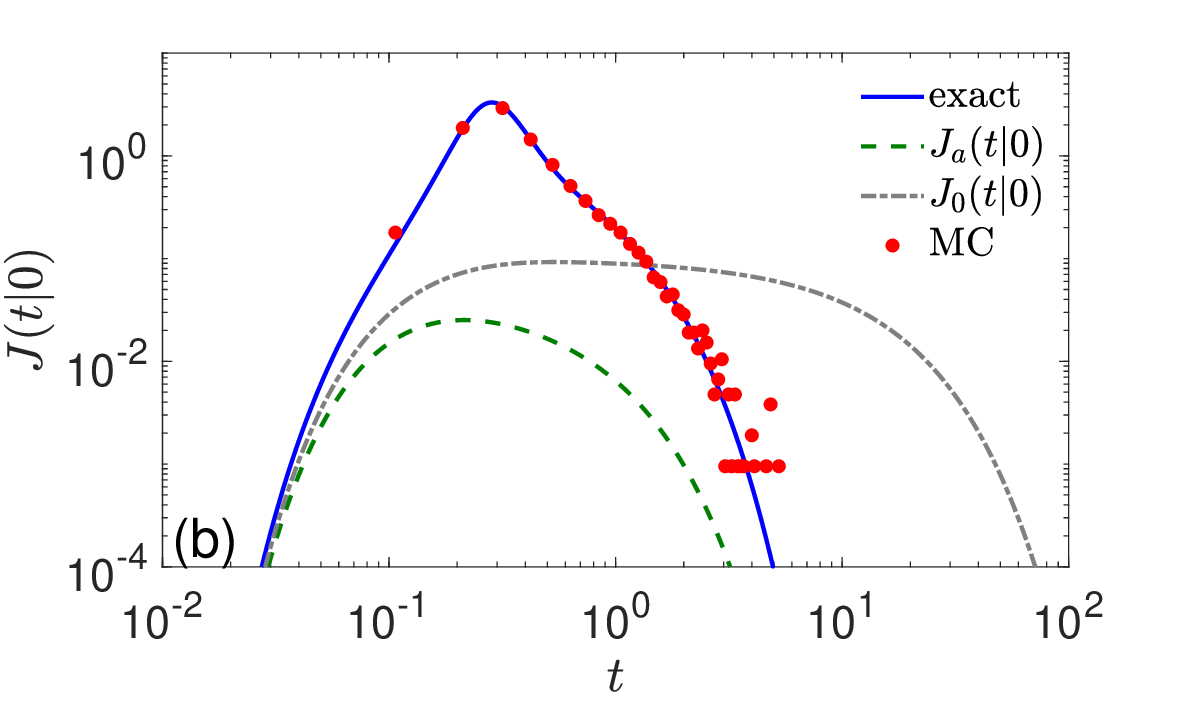} 
\end{center}
\caption{
Survival probability $S(t|0)$ [panel (a)] and the probability density
$J(t|0)$ of the FPT [panel (b)] for the unit interval ($L = 1$), with
$D = 1$, $\qa = 0.1$, and $\qc = 5$.  Solid line presents the exact
quantities obtained via a numerical solution of the integral equation
(\ref{eq:S_integral2}).  Dashed and dash-dotted lines present
$S_{\a}(t|0)$ and $S_0(t|0)$ given by truncated spectral expansions.
Filled circles show the results of Monte Carlo simulations with $10^4$
realizations and the discretization step $a = 0.005$. }
\label{fig:SJ_q5_qa01}
\end{figure}

For comparison, Fig. \ref{fig:SJ_q5_qa01} shows the survival
probability and the probability density for a weakly reactive target
($\qa = 0.1$).  In this case, the conventional search without
autocatalytic dynamics (i.e., for $\qc = 0$) would take much longer,
due to numerous reflections on the target, so that the autocatalytic
dynamics can significantly speed up the search process.  This is
clearly seen on panel (b), where the probability density $J(t|0)$ is
much narrower and centered around shorter times than $J_0(t|0)$.

\subsection{Mean FRT}

After the visual inspection of the survival probabilities in
Sec. \ref{sec:int_survival}, we focus on the analysis of the MFRT.  To
check the quality of the two-sided bound (\ref{eq:T_bounds}), we
recall that the MFRT $T_{\a}(x_0)$ has a simple quadratic form:
\begin{equation} \label{eq:Ta_interval}
T_{\a}(x_0) = - \frac{x_0^2}{2D} + \frac{L(2+ \qa L) (1 + \qc x_0)}{2D (\qa + \qc + \qa \qc L)} \,.
\end{equation}  
Expectedly, $T_{\a}(x_0)$ is a monotonously decreasing function of
both $\qa$ and $\qc$.  In particular, its infimum with respect to
$\qc$ corresponds to the limit $\qc \to
\infty$:
\begin{equation} 
\lim\limits_{\qc\to \infty} T_{\a}(x_0) = T_{\rm min}(x_0) = \frac{x_0}{2D} \biggl(L - x_0 + \frac{L}{1+ \qa L}\biggr).
\end{equation}  
As a consequence, if $x_0 > 0$, even an extremely fast growth of the
population in the limit $\qc \to \infty$ cannot diminish the MFRT
$T(x_0)$ below this value.  This is expected: in order to trigger
cloning events, the first particle has to reach the catalytic region.

Let us now determine the dependence of $T(x_0)$ on the starting point
$x_0$.  For this purpose, we first note that setting $\qc = 0$ to
Eq. (\ref{eq:Ta_interval}) yields
\begin{equation} \label{eq:T0_interval}
T_0(x_0) = \frac{L^2 - x_0^2}{2D} + \frac{L}{D \qa} \,. 
\end{equation}  
In addition, the Green's function for the interval $(0,L)$ is
\begin{align}
& D\tilde{P}(x,0|x_0) = \frac{1}{\qc + \qa + \qc \qa L} \\  \nonumber
& \times \begin{cases} (1+ \qc x)(1 + \qa(L-x_0)) \quad (0 \leq x \leq x_0 \leq L), \cr
(1+ \qc x_0)(1 + \qa(L-x)) \quad (0 \leq x_0 \leq x \leq L). \end{cases}
\end{align}
Setting $x = 0$ yields
\begin{equation}  \label{eq:Pp0}
D\tilde{P}(0,0|x_0) = \frac{1 + \qa (L-x_0)}{\qc + \qa + \qa \qc L} \,,
\end{equation}
whereas $\qc = 0$ implies
\begin{equation}  \label{eq:Pp00}
D\tilde{P}_0(0,0|x_0) = L-x_0 + \frac{1}{\qa }\,.
\end{equation}
As a consequence, Eq. (\ref{eq:T_integral2}) for the MFRT becomes
\begin{equation}
T(x_0) = T_0(x_0) - \qc (L - x_0 + 1/\qa) \bigl[T(0) - T^{(2)}(0) \bigr],
\end{equation}
where $T^{(2)}(0)$ was defined in Eq. (\ref{eq:T2_def}).  In
particular, one has at $x_0 = 0$:
\begin{equation}
T(0) = \frac{T_0(0) + \qc (L + 1/\qa) T^{(2)}(0)}{1 + \qc (L + 1/\qa)} \,,
\end{equation}
where $T_0(0) = L^2/(2D) + L/(D\qa)$.  Expressing the last term via
$T(0)$, one gets
\begin{equation}
T(x_0) = T(0) \biggl(1 - \frac{x_0}{L + 1/\qa}\biggr) + \frac{x_0 T_0(0)}{L + 1/\qa} - \frac{x_0^2}{2D} \,.
\end{equation}
We observe that the MFRT at any point $x_0$ can be easily determined
from $T(0)$.

The explicit quadratic dependence of the MFRT on $x_0$ has immediate
consequences on the search optimality.  Since $d^2 T(x_0)/dx_0^2 =
-1/D < 0$, the function $T(x_0)$ reaches a maximum at $x_0^* =
D(T_0(0) -T(0))/(L + 1/\qa)$.  Using the explicit form of $T_0(0)$, it
is easy to check that $x_0$ lies in the interval $(0,L)$, i.e.,
starting the particle from $x_0^*$ indeed maximizes the MFRT.  In
turn, the minimum is reached at either of the endpoints: $0$ or $L$.
Intuitively, one may expect that placing the starting point on the
target would always be an optimal solution.  However, this is not true
for weakly reactive targets.  Indeed, setting $x_0 = L$, we find
\begin{equation}
T(L) = \frac{1}{1 + \qa L} \biggl(T(0) + \frac{L^2}{2D}\biggr).
\end{equation}
The condition $T(L) \geq T(0)$ is equivalent to $T(0) \leq L/(2\qa
D)$.  When the catalytic rate $\qc$ grows to infinity, the MFRT $T(0)$
vanishes (see Sec. \ref{sec:qc_regimes}).  As a consequence, for any
partially reactive target (i.e., $\qa < \infty$), the above inequality
can be satisfied for a sufficiently large $\qc$.  In other words,
starting on the catalytic region is more advantageous than starting on
the target, if the catalytic rate is large enough.  This explicit
result illustrates the optimality dilemma formulated in
Sec. \ref{sec:intro}: if the target is partially reactive, the
particle may need to return to it a number of times before the
successful reaction; in this situation, it may be more efficient to
cross the interval, to spend some time to produce many clones, and
then to trigger the reaction on the target by a large swarm.  This
conclusion evidently fails for a perfectly reactive target with $\qa =
\infty$.

In the following, we focus on $T(0)$ that incorporates the dependence
on the catalytic rate $\qc$.  This MFRT is computed numerically by
integrating $S(t|0)$ over $t$ (see Eq. (\ref{eq:Tmean_def}) and
Appendix \ref{sec:numerics}).  
When the catalytic rate $\qc$ is small, one can use the perturbative
approach described in Sec. \ref{sec:qc_regimes}; in particular,
Eq. (\ref{eq:T_qcsmall}) gives the two-term approximation:
%
%
\begin{equation}
T(0) \approx T_0(0) - \qc (L + 1/\qa) \biggl\{T_0(0) - \int\limits_0^\infty dt\, S_0^2(t|0) \biggr\} + O(q_{\rm c}^2),
\end{equation}
where we used Eq. (\ref{eq:Pp00}).  Higher-order corrections can also
be evaluated in a systematic way.  
In turn, an accurate evaluation of the large-$\qc$ asymptotic behavior
is much more challenging and presents an open problem even for the
interval (see Sec. \ref{sec:discussion}).

\begin{figure}
\begin{center}
\includegraphics[width=88mm]{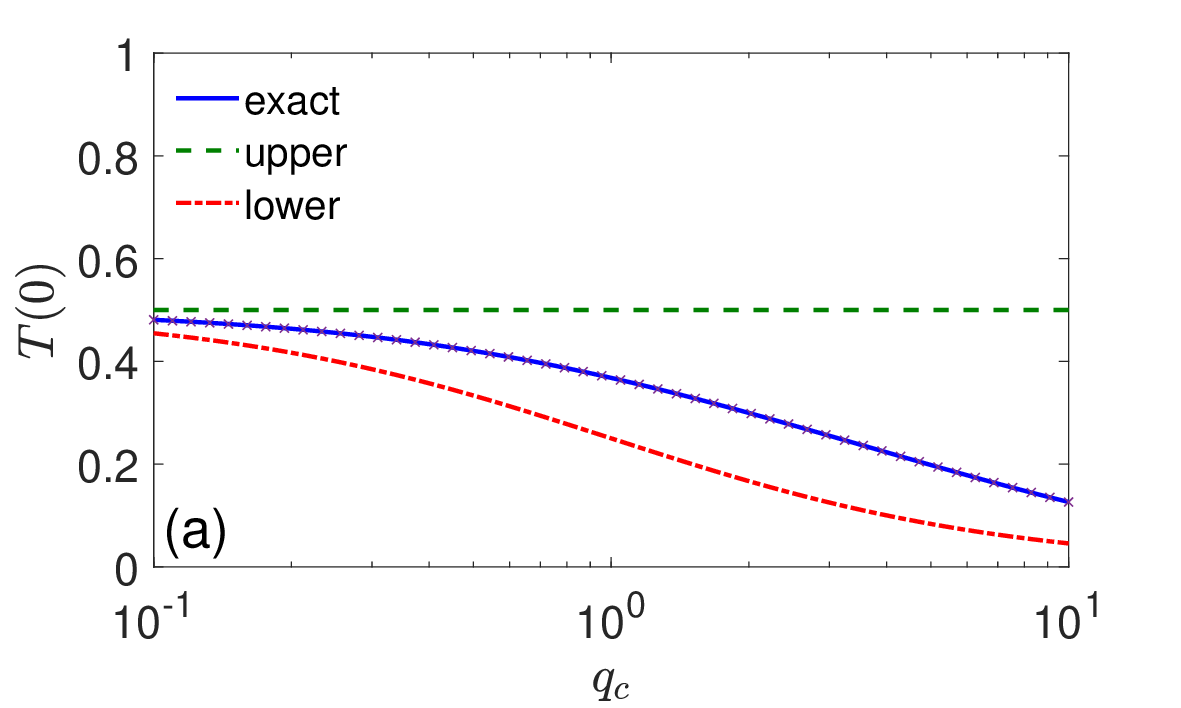} 
\includegraphics[width=88mm]{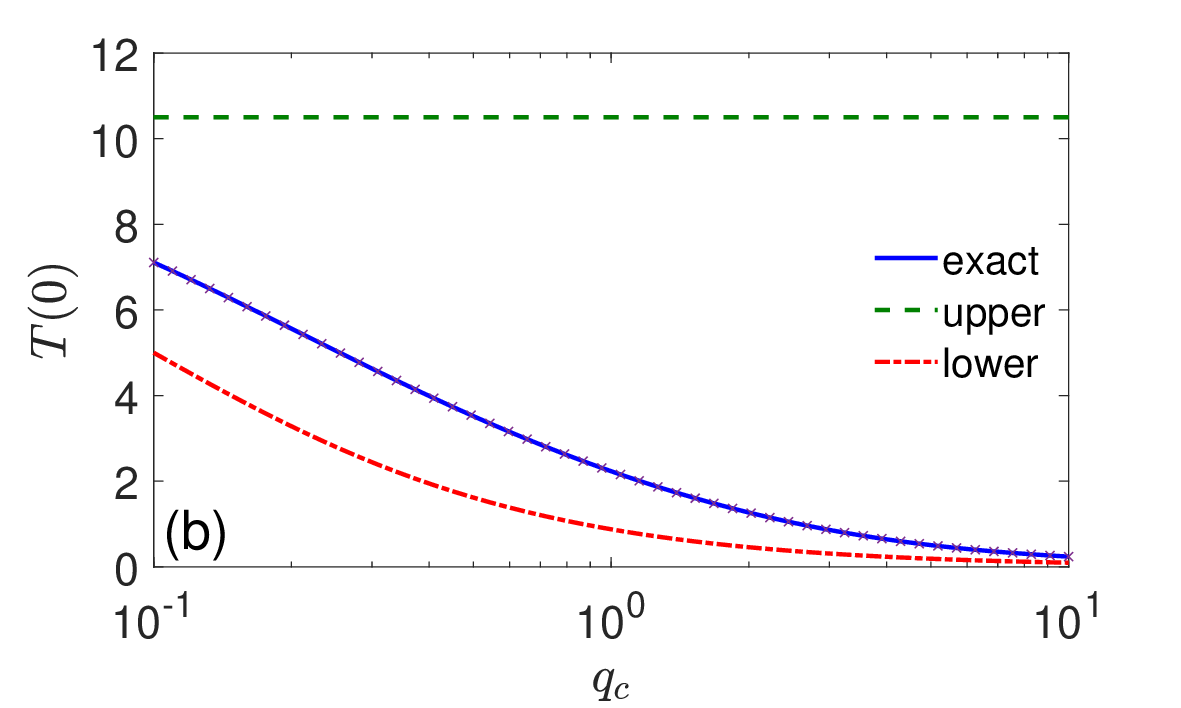} 
\end{center}
\caption{
The MFRT $T(0)$ as a function of the catalytic rate $\qc$ for the unit
interval ($L = 1$), with $D = 1$, $\qa = \infty$ [panel (a)] and $\qa
= 0.1$ [panel (b)].  Solid line presents the exact solution obtained
via a numerical integration of the survival probability $S(t|0)$ via
Eq. (\ref{eq:Tmean_def}) (crosses indicate results obtained with a
timestep reduced by a factor of two, confirming numerical
convergence).  Dashed and dash-dotted lines present the upper and
lower bounds $T_0(0)$ and $T_{\a}(0)$ given by
Eqs. (\ref{eq:T0_interval},
\ref{eq:Ta_interval}).}
\label{fig:Tmean2}
\end{figure}

Figure \ref{fig:Tmean2} presents the MFRT $T(0)$ as a function of the
catalytic rate $\qc$.  When the target is perfectly reactive (panel
(a), $\qa = \infty$), the autocatalytic dynamics with moderate $\qc$
provides only a minor reduction of the MFRT as compared to the
conventional search ($\qc = 0$): e.g., even at $\qc = 1$, we have
$T(0) \approx 0.37$, whereas $T_0(0) = 0.5$.  In turn, the search
efficiency is improved as $\qc$ increases; for instance, at $\qc =
10$, one achieves a four-fold reduction of the MFRT, from $T_0(0) =
0.5$ to $T(0) \approx 0.13$.

The situation changes dramatically for weakly reactive targets (panel
(b), $\qa = 0.1$).  Because successful reaction typically requires
multiple returns to the target, the MFRT $T_0(0)$ in the conventional
search is large: $T_0(0) = 10.5$ according to
Eq. (\ref{eq:T0_interval}).  In this scenario, cloning of particles on
$\Gammac$ can significantly improve the search.  For instance, at $\qc
= 1$, we find $T(0) \approx 2.23$, i.e., almost a five-fold reduction
as compared to $T_0(0)$.  The improvement becomes even more pronounced
as the catalytic rate increases.  We conclude that the autocatalytic
dynamics is more advantageous for finding targets with low reactivity.
This result highlights that autocatalytic replication is particularly
effective when repeated unsuccessful encounters dominate the search
process.

\begin{figure}
\begin{center}
\includegraphics[width=88mm]{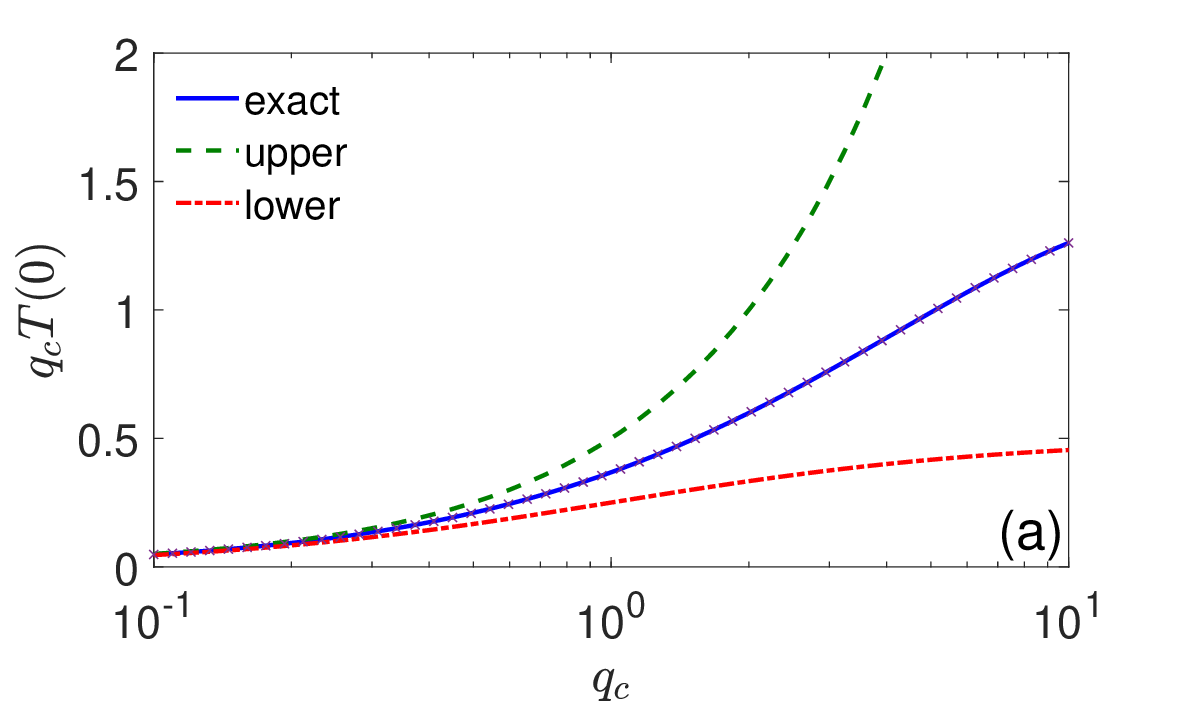} 
\includegraphics[width=88mm]{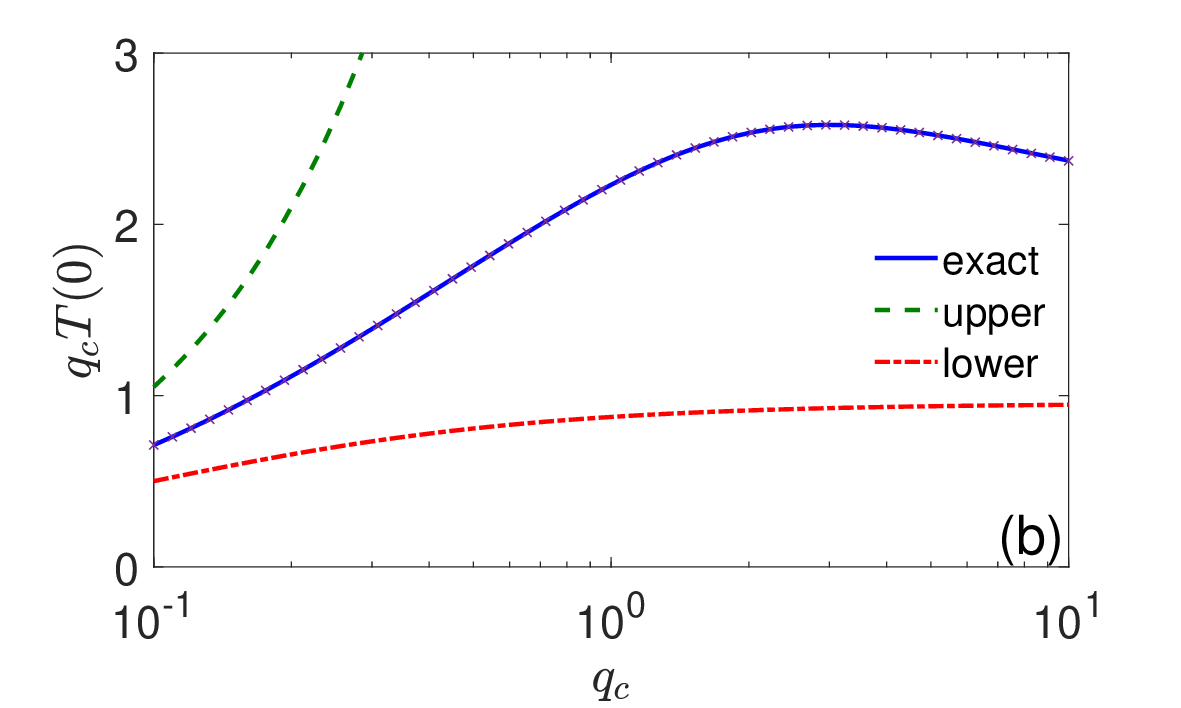} 
\end{center}
\caption{
The MFRT $T(0)$ (multiplied by $\qc$) as a function of the catalytic
rate $\qc$ for the unit interval ($L = 1$), with $D = 1$, $\qa =
\infty$ [panel (a)] and $\qa = 0.1$ [panel (b)].  Solid line presents
the exact solution obtained via a numerical integration of the
survival probability $S(t|0)$ via Eq. (\ref{eq:Tmean_def}) (crosses
indicate results obtained with a timestep reduced by a factor of two,
confirming numerical convergence).  Dashed and dash-dotted lines
present the upper and lower bounds $\qc T_0(0)$ and $\qc T_{\a}(0)$
given by Eqs. (\ref{eq:T0_interval},
\ref{eq:Ta_interval}).}
\label{fig:Tmean3}
\end{figure}

How efficient the autocatalytic dynamics can be at large $\qc$?  Since
$T_{\a}(0)$ is the lower bound, the search improvement cannot be
better than
\begin{equation}
\frac{T_\a(0)}{T_0(0)} = \frac{1}{1 + \qc L + \qc/\qa} \,.
\end{equation}
For a perfectly reactive target, the search improvement is limited by
$1/(1 + \qc L)$ but it can further be enhanced for weakly reactive
targets.  In both cases, the reduction scales as $1/\qc$ at large
$\qc$.  To reveal the asymptotic behavior more clearly, we therefore
rescale the MFRT by $\qc$.  On both panels of Fig. \ref{fig:Tmean3},
$\qc T(0)$ at $\qc = 10$ is around $2.5 - 2.8$ times larger than $\qa
T_\a(0)$.  This numerical evidence suggests that $T(0)$ scales as
$1/\qc$ at large $\qc$, even though finding the precise form of this
asymptotic behavior remains an interesting open analytical problem
(see Sec. \ref{sec:discussion}).

\section{Discussion and conclusion}
\label{sec:discussion}

The present work extends classical first-passage theory by
incorporating autocatalytic branching events localized on the
boundary.  
Branching processes have many applications in disciplines as diverse
as physics, chemistry, biology, social sciences, and risk management
\cite{Harris,Athreya,Williams,Kimmel,Murrey}.  Branching events in the
bulk \cite{Kolmogorov47,Sevastyanov58,Skorohod64} naturally lead to
nonlinear PDEs, including the seminal Fisher-KKP equation
\cite{Fisher37,Kolmogorov37} and many other nonlinear
reaction-diffusion equations \cite{Grindrod}.  Over the past three
decades, significant efforts were dedicated to develop a rigorous
mathematical description of branching events on lower-dimensional
subsets in the continuous setting
\cite{LeGall,Dynkin,DelGross76,Dawson99,Klenke00,Englander04,Delmas05,Morters05,Englander07,Bocharov14}
and on a subset of catalytic sites for lattice random walks
\cite{Redner84,Albeverio98,Bogachev98,Kesten03,Vatutin04,Vatutin07,Hu12,Topchi13,Bulinskaya18,Bauer21}.
In particular, the opposite effects of branching and annihilation
events onto the population dynamics were thoroughly analyzed (see,
e.g, \cite{Canet04a,Canet04b,Odor04,Grebenkov26a} and reference
therein).  In turn, the first-passage statistics within the realm of
autocatalytic dynamics were not studied in depth, to our knowledge.

We investigated diffusion inside a bounded domain whose boundary
contains (up to) three distinct regions: a partially reactive target
$\Gammaa$, a catalytic region $\Gammac$, and the remaining reflecting
part $\Gammar$.  When there is no catalytic region (i.e., $\Gammac =
\emptyset$), finding the survival probability and the distribution of
the associated first-reaction time is the classical problem in the
theory of diffusion-controlled reactions
\cite{Redner,Schuss,Metzler,Masoliver,Lindenberg,Grebenkov,Grebenkov23n}.  
Because the catalytic region generates additional searchers through
replication, its presence is expected to accelerate the search
process.  The central quantity of our study was the first-reaction
time $\T$ on the target by any of the particles in the population.  We
derived two equivalent representations -- the integral equation
(\ref{eq:S_integral}) and the PDE (\ref{eq:S_PDE}) -- for the survival
probability $S(t|\x_0) = \P_{\x_0}\{ \T > t\}$ that fully
characterizes the random variable $\T$.  Accounting for cloning events
on the catalytic region $\Gammac$ led to a nonlinear term in the
integral equation and to a nonlinear Robin-type boundary condition in
PDE.  In particular, even though the mean first-reaction time
$T(\x_0)$ satisfies the Poisson equation (\ref{eq:MFPT_PDE}), as in
the conventional setting, this PDE problem is not self-contained,
given that the boundary condition (\ref{eq:MFPT_qc}) includes the
integral of the squared survival probability.  These nonlinear
features were the major challenges in the analysis of the
first-reaction time.  Nevertheless, we managed to obtain the lower and
the upper bounds for the survival probability $S(t|\x_0)$ and,
consequently, for all the positive-order moments of $\T$.  We also
deduced the long-time behavior of the survival probability and
explained that the critical catalytic rate governs the long-time
population dynamics but not the statistics of the first successful
reaction.  The regimes of both low and high catalytic rates were
briefly discussed.  Our results therefore demonstrate that localized
autocatalytic replication fundamentally alters first-passage
statistics without changing the underlying diffusion operator.

In order to get a deeper insight onto the dependence of the FRT on the
autocatalytic dynamics, we focused on a simple yet representative
example of the interval $(0,L)$, whose two endpoints were the
catalytic region and the target.  We developed a numerical scheme for
solving the integral equation to access the survival probability
$S(t|x_0)$, the probability density of the FRT, and its mean value
$T(x_0)$.  In particular, we further showed how $T(x_0)$ can be
expressed in terms of $T(0)$ and then analyzed the optimality of the
MFRT; in particular, we argued that starting from the target is not
necessarily the optimal strategy when the reaction rate $\qa$ is
finite.  This observation may be particularly relevant for
diffusion-controlled biochemical reactions, where reaction
probabilities are often much smaller than unity.

Our main focus was on the dependence of $T(0)$ on two parameters: the
catalytic rate $\qc$ and the reaction rate $\qa$.  For perfectly (or
highly) reactive targets (with infinite or large $\qa$), the effect of
autocatalytic dynamics with small or moderate $\qc$ was minor; indeed,
the MFRT $T(0)$ could be significantly reduced only at large $\qc$.
The situation was drastically different for weakly reactive targets
(small $\qa$).  Here, even a weak autocatalytic dynamics could
significantly decrease $T(0)$.  We anticipate that our conclusions
that were obtained for the interval, remain qualitatively valid in
more general settings.  For instance, an extension to concentric
circular annuli and spherical shells is rather straightforward due to
their rotational invariance.  Moreover, as most chemical and
biophysical applications deal with small targets, the cloning
mechanism can present a considerable advantage, as in our setting of a
weakly reactive target.  Further investigation of this small-target
limit presents a promising perspective of this work.

Another important open problem concerns the asymptotic behavior of the
survival probability and the MFRT at large catalytic rate $\qc$.  Even
in the case of an interval, for which the ``ingredients'' $S_0(t|0)$
and $P_0(0,t|0)$ determining the integral equation
(\ref{eq:Sinterval_eq}) are explicitly known, this problem turned out
to be surprisingly difficult.  (i) From the population dynamics
perspective, the mean number of particles grows exponentially fast,
approximately as $e^{q_{\rm c}^2 Dt}$, that prohibits Monte Carlo
simulations.  In this regime, it may be instructive to compare the
autocatalytic search to the large-$N$ asymptotic behavior of the
extremal statistics for a fixed-size population (with $N$ particles).
(ii) As we argued in Appendix \ref{sec:Ashort}, $S(t|0) \approx
S_0(t|0) = 1 + O(e^{-C/t})$ at very short times; as a consequence, one
can linearize the integral equation for $u(t) = 1-S(t|0)$ but its
solution exhibits a steep exponential growth, $\propto e^{q_{\rm c}^2
Dt}$, until the nonlinear term becomes relevant.  As a consequence, an
accurate numerical solution of the nonlinear equation
(\ref{eq:Sinterval_eq}) with large $\qc$ via standard quadratures is
computationally demanding due to the need of using extremely small
timesteps, whereas the numerical results are sensitive to round-off
errors.
(iii) At short times, one has $P_0(0,t|0) \approx 1/\sqrt{\pi Dt}$, so
that the integral term is close to the fractional integral operator of
order $1/2$.  Its formal inversion allows one to recast the original
problem at short times as a forced fractional logistic problem:
\begin{equation}  \label{eq:frac_logistic}
{\mathcal D}_t^{\frac12} u = f + q u (1-u),
\end{equation}
where $q \propto \qc$, the forcing term $f(t)$ is related to
$S_0(t|0)$, and ${\mathcal D}_t^{\frac12}$ is the fractional
derivative of order $1/2$.  Equation (\ref{eq:frac_logistic}) reveals
an unexpected connection between autocatalytic search and nonlinear
fractional dynamical systems.  In contrast to the classical setting of
forced logistic equation, which is integrable, the solution of the
fractional logistic equation (\ref{eq:frac_logistic}) depends on the
entire history, i.e., it is effectively an infinite-dimensional
dynamical system, which may exhibit rich dynamical behavior,
especially for large $q$.  Further analysis of the asymptotic behavior
at large $\qc$ is an interesting open problem.

Finally, we stress that the considered model of the autocatalytic
dynamics can naturally be generalized in different ways: (i) If each
branching event produces $m$ particles, the term $S^2(t|\x)$ in the
integral equation (\ref{eq:S_integral}) is replaced by $S^m(t|\x)$;
moreover, one can randomly pick the branching order $\hat{m}$ from a
given distribution, in which case one deals with the expectation $\E\{
S^{\hat{m}}(t|\x)\}$.  (ii) In many applications, the particles have a
finite lifetime (e.g., radioactive decay), or diffuse in a reactive
medium that can spontaneously destroy the particle
\cite{Yuste13,Meerson15,Grebenkov17}; such ``mortal'' particles may
fail to find the target due to their death or passivation; the
inclusion of a finite lifetime with a rate $\nu$ can be realized by
adding the term $-\nu S$ to the right-hand side of the diffusion
equation (\ref{eq:S_diff}).  (iii) Throughout this work, we assumed
that the search was initiated with a single particle; as the
conventional diffusion-controlled reactions are described by linear
equations, a passage from a single initial particle to a configuration
with many particles is straightforward; for instance, the distribution
of the fastest FRT $\T_N$ for $N$ independent particles started from
$\x_0$ is fully determined as $\P_{\x_0}\{ \T_N > t\} =
[S_0(t|\x_0)]^N$, in terms of the survival probability $S_0(t|\x_0)$
for a single particle; in sharp contrast, the nonlinear nature of
branching events makes the impact of multiple initial particles highly
sophisticated.  (iv) Finally, the ordinary diffusion with constant
diffusivity $D$ can be replaced by a more general diffusion process in
a heterogeneous medium or in the presence of an external potential, as
well as anomalous diffusions (like the fractional diffusion equation).
All these extensions open several promising research directions.
More broadly, the present work establishes a framework for
incorporating localized branching mechanisms into first-passage
theory.  We expect that similar ideas will prove useful in stochastic
population dynamics, intracellular transport, catalytic reactions, and
search processes involving self-replicating agents.

\begin{acknowledgments}
The author thanks Mr. Y. Ye for fruitful discussions.  The author
acknowledges the Simons Foundation for supporting his sabbatical
sojourn in 2024 at the CRM (University of Montr\'eal, Canada), as well
as the Alexander von Humboldt Foundation for support within a Bessel
Prize award.
\end{acknowledgments}

\appendix
\section{Numerical solution}
\label{sec:numerics}

This appendix summarizes the mathematical ingredients used in the
numerical analysis of autocatalytic reactions and related FRTs on the
interval $(0,L)$ with the catalytic endpoint $\Gammac = \{0\}$ and the
reactive endpoint $\Gammaa = \{L\}$: the Laplacian eigenbasis and the
propagator (Sec. \ref{sec:int_Laplacian}), a numerical method for
solving integral equations (Sec. \ref{sec:num_integral}), and an
independent Monte Carlo technique (Sec. \ref{sec:MC}).

\subsection{Spectral representations}
\label{sec:int_Laplacian}

The Laplacian eigenvalues and eigenfunctions are
\begin{subequations}  \label{eq:uk}
\begin{align}
\lambda_k & = \alpha_k^2/L^2,  \\
u_k(x) & = \sqrt{\frac{2}{L}} \beta_k \biggl(\cos(\alpha_k x/L) + \frac{h_1}{\alpha_k} \sin(\alpha_k x/L)\biggr),
\end{align}
\end{subequations}
where $h_1 = \qc L$, $h_2 = \qa L$, 
\begin{equation}  \label{eq:betak}
\beta_k = \biggl(\frac{\alpha_k^2}{\alpha_k^2 + h_1 + h_1^2 + h_2(\alpha_k^2 + h_1^2)/(\alpha_k^2 + h_2^2)}\biggr)^{1/2} 
\end{equation}
are the normalization constants, and $\alpha_k$ are the positive
solutions of the equation:
\begin{equation}  \label{eq:alphak}
\frac{\tan(\alpha_k)}{\alpha_k} = \frac{h_1 + h_2}{\alpha_k^2 - h_1 h_2} \qquad (k = 0,1,2,\ldots).
\end{equation}
The integral of an eigenfunction over the interval reads
\begin{equation}
\int\limits_0^L dx \, u_k(x) = \sqrt{2L} \frac{\beta_k}{\alpha_k^2} \left(h_1 + 
(-1)^k h_2 \sqrt{\frac{\alpha_k^2 + h_1^2}{\alpha_k^2 + h_2^2}}\right).
\end{equation}
In the special case of a perfectly reactive target ($\qa = \infty$),
Eqs. (\ref{eq:betak}, \ref{eq:alphak}) are reduced to
\begin{equation}
\beta_k = \biggl(\frac{\alpha_k^2}{\alpha_k^2 + h_1 + h_1^2}\biggr)^{1/2} 
\end{equation}
and
\begin{equation}  \label{eq:alphak_qaInf}
\frac{\tan(\alpha_k)}{\alpha_k} = -\frac{1}{h_1} \,.
\end{equation}

These eigenmodes provide spectral representations of the propagator
$P(x,t|x_0)$ and the survival probability $S_{\a}(t|x_0)$ via their
spectral expansions; for instance,
\begin{align*}
S_\a(t|x_0) = \sum\limits_{k=0}^\infty e^{-Dt\lambda_k} u_k(x_0) \int\limits_0^L dx \, u_k(x).
\end{align*}
Truncating these expansions after $K$ terms allows one to accurately
compute them for any $t > \delta$ with $\delta$ such that
$D\delta\lambda_K \gg 1$.  For $x = x_0 = 0$, we have the following
short-time behavior, which follows from the exact propagator on the
half-line:
\begin{equation}  \label{eq:P_asympt}
P(0,t|0) \approx \frac{1}{\sqrt{\pi Dt}} - \qc \erfcx\bigl(\qc \sqrt{Dt}\bigr) \quad (t\to 0),
\end{equation}
with exponentially small corrections, where the last term is actually
the next-order correction, given that $\erfcx(z) \approx 1 + O(z)$ as
$z\to 0$ (here $\erfcx(z) = e^{z^2}\erfc(z)$ is the scaled
complementary error function).  

The Laplace transforms of $P(x,t|x_0)$ and $S_{\a}(t|x_0)$ admit a
fully explicit form.  In particular, one has
\begin{equation}  \label{eq:tildeP}
D\tilde{P}(0,p|0) = \frac{\qa \sinh(\alpha L) + \alpha \cosh(\alpha L)}
{(\alpha^2 + \qa \qc)\sinh(\alpha L) + \alpha (\qa + \qc)\cosh(\alpha L)} 
\end{equation}
and
\begin{equation}
\tilde{S}_{\a}(p|0) = \frac{[\alpha \sinh(\alpha L) + \qa (\cosh(\alpha L)-1)]/(D\alpha )}
{(\alpha^2 + \qa \qc)\sinh(\alpha L) + \alpha (\qa + \qc)\cosh(\alpha L)}  \,,
\end{equation}
where $\alpha = \sqrt{p/D}$.  In turn, $D\tilde{P}_0(0,p|0)$ and
$\tilde{S}_0(p|0)$ follow from these expressions by setting $\qc = 0$.
In the limit $p\to \infty$, we have then
\begin{equation}
D\tilde{P}_0(0,p|0) \simeq \frac{1}{\alpha} + \frac{2}{\alpha} \, \frac{\alpha - \qa}{\alpha + \qa} e^{-2\alpha L} + O(e^{-4\alpha L}).
\end{equation}
For a finite $\qa$, the ratio $(\alpha - \qa)/(\alpha + \qa)$ is close
to $1$ as $p\to \infty$, from which the short-time behavior follows:
\begin{equation}  \label{eq:P0_short}
P_0(0,t|0) \simeq \frac{1}{\sqrt{\pi Dt}} \bigl(1 + 2 e^{-L^2/(Dt)} + \cdots\bigr)
\end{equation}
(note that these two leading-order terms are independent of $\qa$).
Similarly, for a finite $\qa$, one has
\begin{equation}
\tilde{S}_0(p|0) \simeq \frac{1- 2e^{-\alpha L}}{p}  + O(e^{-2\alpha L}),
\end{equation}
and the Laplace transform inversion yields the short-time behavior
\begin{equation}  \label{eq:S0_short}
S_0(t|0) \approx 1 - \frac{4\sqrt{Dt}}{\sqrt{\pi} L} e^{-L^2/(4Dt)} . 
\end{equation}

In the special case $\qc = 0$ and $\qa = \infty$, the spectrum is
fully explicit:
\begin{subequations}
\begin{align}
\lambda_k & = \pi^2 (k+1/2)^2/L^2, \\
u_k(x) & = \sqrt{2/L} \cos(\pi (k+1/2)x/L),
\end{align}
\end{subequations}
so that
\begin{subequations}  \label{eq:P0S0_v1}
\begin{align}
P_0(0,t|0) & = \frac{2}{L}\sum\limits_{k=0}^\infty e^{-\pi^2(k+1/2)^2 Dt/L^2} , \\
S_0(t|0) & = 2\sum\limits_{k=0}^\infty (-1)^k \frac{e^{-\pi^2(k+1/2)^2 Dt/L^2}}{\pi(k+1/2)} \, . 
\end{align}
\end{subequations}
Using the Poisson summation formula, one also gets equivalent
representations, which are more suitable at short times:
\begin{subequations}  \label{eq:P0S0_v2}
\begin{align}  
P_0(t|0) & = \frac{1}{\sqrt{\pi Dt}}\biggl(1 + 2\sum\limits_{k=1}^\infty (-1)^k e^{-k^2 L^2/(Dt)}\biggr), \\ 
S_0(t|0) & = 1 - 2\sum\limits_{k=0}^\infty (-1)^k \erfc((k+1/2)L/\sqrt{Dt})  . 
\end{align}
\end{subequations}
In particular, one recovers the short-time behavior
(\ref{eq:S0_short}).

\subsection{Quadrature solver}
\label{sec:num_integral}

We need to solve numerically the nonlinear equation
(\ref{eq:Sinterval_eq}) with $x_0 = 0$.  Following
\cite{Grebenkov26c}, we introduce
\begin{equation}
p(t) = \qc D P_0(0,t|0) \, \sqrt{t}
\end{equation}
to remove the weak square-root singularity of the propagator
$P_0(0,t|0)$ at short times.  In fact, according to
Eq. (\ref{eq:P_asympt}), we have $p(0) = \qc \sqrt{D}/\sqrt{\pi}$.
Discretizing the time interval $(0,t)$ into $k$ subintervals of
duration $\delta = t/k$, we get for $x_0 = 0$
\begin{equation}
S(k\delta|0) \approx S_0(k\delta|0) + \sum\limits_{j=0}^k w_j \bigl[S^2((k-j)\delta|0) - S((k-j)\delta|0)\bigr],
\end{equation}
with the quadrature weights $w_j$:
\begin{subequations}  \label{eq:wj}
\begin{align}
w_j & = \frac{a_j + a_{j-1}}{2}  \qquad (j=0,1,\ldots,k),  \\
a_j & = \bigl(\sqrt{t_{j+1}} - \sqrt{t_j}\bigr) \bigl[p(j\delta) + p((j+1)\delta)\bigr]  ,
\end{align}
\end{subequations}
and $a_{-1} = a_k = 0$ (see more details in \cite{Grebenkov26c}).  In
the shown numerical examples, the time step was set to $\delta =
10^{-3}$.
Isolating the $j = 0$ contribution yields the quadratic equation,
whose solution is
\begin{subequations}
\begin{align}
S(k\delta|0) & \approx \frac{2f_k}{1+w_0 + \sqrt{(1+w_0)^2 -4 w_0 f_k}} \,,  \\
f_k & = S_{0}(k\delta|0) + \sum\limits_{j=1}^k w_j S^2((k-j)\delta|0).
\end{align}
\end{subequations}
Once the survival probability $S(t|0)$ has been computed, the solution
for any starting position $x_0 >0$ follows directly from
Eq. (\ref{eq:Sinterval_eq}).  In addition, the mean FRT can be
approximated as
\begin{equation}
T(0) \approx \delta \sum\limits_{k=0}^K S(k\delta|0),
\end{equation}
with a large enough $K$.  Numerically, the summation is continued
until the survival probability falls below a prescribed tolerance.
(e.g., $10^{-4}$).  For the computation of $T(0)$ shown in
Figs. \ref{fig:Tmean2} and \ref{fig:Tmean3}, we first determined
$t_{\rm max}$ such that $S_\a(t|0) < 10^{-4}$ for $t > t_{\rm max}$,
and then set $\delta = t_{\rm max}/10^4$.  This adaptive procedure
naturally employs smaller timesteps for larger catalytic rates.
Numerical convergence was verified by repeating the calculations with
a timestep reduced by a factor of two (results shown by crosses).

In turn, the integral equation (\ref{eq:J_integral2}) is {\it linear}
with respect to $J(t|x_0)$ and can thus be solved by standard
deconvolution techniques (e.g., via a discrete FFT transformation).
However, we keep using the above direct scheme even for this equation.
Here, we simply write
\begin{align*}
J(k\delta|0) & \approx J_0(k\delta|0) \\
& + \sum\limits_{j=0}^k w_j \bigl[2S((k-j)\delta|0) - 1\bigr] J((k-j)\delta|0)),
\end{align*}
from which
\begin{align} 
J(k\delta|0) & \approx \frac{1}{1 + w_0(1 - 2S(k\delta|0))}  \biggl[J_0(k\delta|0) \\  \nonumber
& + \sum\limits_{j=1}^k w_j (2S((k-j)\delta|0) - 1) J((k-j)\delta|0))\biggr].
\end{align}
In practice, we use this scheme at short times; in turn, at longer
times, it is easier and more stable to evaluate $J(k\delta|0)$ from a
finite-difference approximation: 
\begin{equation}
J(k\delta|0) \approx \frac{S(k\delta|0) - S((k-1)\delta|0)}{\delta} \,.
\end{equation}

\subsection{Monte Carlo simulations}
\label{sec:MC}

For validation purposes, we also perform independent Monte Carlo
simulations of a random walk on the interval $(0,L)$.  For this
purpose, we discretize the interval with a step $a$ and introduce the
associated timestep $\delta = a^2/(2D)$ (we set $a = 0.005$).  When
the particle is on the catalytic site ($x = 0$), it can either jump to
the neighboring site $a$ with probability $1 - p_{\c}$, or be split
into two particles with the probability $p_{\c} = a \qc/(1 + a \qc)$.
This corresponds to a standard probabilistic interpretation of partial
reactivity $\qc$ with the Robin boundary condition (see, e.g.,
\cite{Grebenkov03}).  Simulations are advanced with the timestep
$\delta$ for all present particles, until at least one particle
reaches the reactive endpoint at $x = L$.  Here, the particle either
reacts with the probability $p_{\a} = a \qa /(1+ a \qa)$ and the
simulation stops, or it moves to the neighboring site $L-a$ to
continue diffusion.  Although more efficient algorithms have been
developed for modeling autocatalytic dynamics (see, e.g.,
\cite{Grebenkov26c}), this basic scheme was sufficient for our
validation purposes.  To get adequate statistical accuracy, the
simulation is repeated $M = 10^4$ times.

\subsection{Spectral ODE solver}

To verify that the numerical solution of the nonlinear integral
equation is independent of the discretization strategy, we also
implemented an alternative solver based on spectral decomposition.
Since both functions $S_\a(t|0)$ and $P(0,t|0)$ admit spectral
representations, one can reduce the nonlinear integral equation to a
system of nonlinear ordinary differential equations (ODEs).  Let us
consider a nonlinear Volterra equation
\begin{equation}
u(t) = f(t) + \int\limits_0^t dt' \, g(t-t') \,F(u(t')),
\end{equation}
where $F(z)$ is a prescribed function, and $f(t)$ and $g(t)$ can be
accurately approximated via truncated exponential expansions:
\begin{equation}
f(t) = \sum\limits_{k=0}^{K} a_k e^{-b_k t}, \quad 
g(t) = \sum\limits_{k=0}^{K} c_k e^{-b_k t}.
\end{equation}
Setting
\begin{equation}  \label{eq:yk_def}
y_k(t) = a_k e^{-b_k t} + c_k \int\limits_0^t dt' \, e^{-b_k(t-t')} \, F(u(t')),
\end{equation}
we have 
\begin{equation}
u(t) = \sum\limits_{k=0}^K y_k(t).
\end{equation}
Evaluating the time derivatives of $y_k(t)$, one gets the system of
nonlinear ODEs:
\begin{equation}
y_k'(t) = -b_k y_k + c_k F(y_0(t) + \cdots + y_K(t)).
\end{equation} 
The resulting ODE system can be integrated using standard adaptive
Runge-Kutta methods.  Here, we solved these equations iteratively with
a small timestep $\delta$ by using Eq. (\ref{eq:yk_def}):
\begin{align*}
y_k(t+\delta) & = a_k e^{-b_k (t+\delta)} + c_k \int\limits_0^{t+\delta} dt' \, e^{-b_k(t+\delta-t')} \, F(u(t')) \\
& = e^{-b_k \delta} y_k(t) + c_k \int\limits_t^{t+\delta} dt' \, e^{-b_k(t+\delta-t')} \, F(u(t')) \\
& \approx e^{-b_k \delta} y_k(t) + c_k \frac{1-e^{-b_k \delta}}{b_k} F(u(t)). 
\end{align*}
This numerical scheme was efficient for solving
Eq. (\ref{eq:Sinterval_eq}), in which $S_0(t|0)$ and $P_0(0,t|0)$
admit easily accessible spectral expansions.  It was accurate for
moderate $\qc$ but it required very small time steps for large $\qc$.
This scheme was only used for validation purposes (results not shown).

\section{Short-time behavior}
\label{sec:Ashort}

In this Appendix, we sketch some preliminary results on the short-time
behavior.  Setting $\bar{S}(t) = 1 - S(t|0)$, we first rewrite
Eq. (\ref{eq:Sinterval_eq}) as
\begin{equation}
\bar{S}(t) = 1 - S_0(t|0) + \qc D\int\limits_0^t dt' \, P_0(0,t-t'|0) \, \bar{S}(t') \bigl(1 - \bar{S}(t')\bigr).
\end{equation}
According to the asymptotic relation (\ref{eq:P0_short},
\ref{eq:S0_short}), one has $P_0(0,t|0) \approx 1/\sqrt{\pi Dt}$,
whereas $1- S_0(t|0)$ is exponentially small at short times.  At short
times, $\bar{S}(t)$ is thus small, so that the nonlinear term can be
neglected, and the approximate linearized equation reads
\begin{equation}
\bar{S}(t) \approx f(t) + \qc \sqrt{D/\pi} \int\limits_0^t \frac{dt'}{\sqrt{t-t'}} \, \bar{S}(t')  ,
\end{equation}
with
\begin{equation}
f(t) = \frac{4\sqrt{Dt}}{\sqrt{\pi} L} e^{-L^2/(4Dt)} 
\end{equation}
representing $1-S_0(t|0)$.  The above linearized equation can be
solved by applying the Laplace transform.  Using $\L\{1/\sqrt{\pi t}\}
= 1/\sqrt{p}$ and
\begin{equation}
\tilde{f}(p) = 2 \frac{e^{-L\sqrt{p/D}}}{p} \biggl(1 + \frac{1}{L \sqrt{p/D}}\biggr),
\end{equation}
we get
\begin{equation}
\tilde{\bar{S}}(p) \approx \frac{2 e^{-L\sqrt{p/D}} (1 + L\sqrt{p/D})}{p^{3/2} L/\sqrt{D} (1 - \qc \sqrt{D/p})} \,.
\end{equation} 
Using the identities
\begin{align}
\L^{-1} \biggl\{ \frac{e^{-a\sqrt{p}}}{p(b + \sqrt{p})}\biggr\} & = \frac{1}{b} \erfc\bigl(a/\sqrt{4t}) \\  \nonumber
& - \frac{1}{b} e^{-a^2/(4t)} \erfcx\bigl(a/\sqrt{4t} + b \sqrt{t}\bigr)
\end{align}
and
\begin{equation}
\L^{-1} \biggl\{ \frac{e^{-a\sqrt{p}}}{b + \sqrt{p}}\biggr\} = e^{-a^2/(4t)} \biggl(\frac{1}{\sqrt{\pi t}} 
- b \, \erfcx\bigl(a/\sqrt{4t} + b \sqrt{t}\bigr)\biggr),
\end{equation}
we find
\begin{align} \nonumber
\bar{S}(t) & \approx \frac{2 e^{-L^2/(4Dt)}}{\qc L} \biggl\{(1 + \qc L)\, \erfcx\biggl(\frac{L}{\sqrt{4Dt}} - \qc \sqrt{Dt}\biggr) \\  
\label{eq:u_asympt_moderate}
& -  \erfcx\bigl(L/\sqrt{4Dt}\bigr) \biggr\}.
\end{align}
We emphasize that this expression is meaningful only when $\bar{S}(t)
\ll 1$, i.e., at short times.  At very short times, one can use
$\erfcx(z) \approx 1/(\sqrt{\pi} z)$ as $z\to 0$ in order to get
\begin{equation}  \label{eq:u_asympt_small}
\bar{S}(t) \simeq f(t) \approx 1 - S_0(t|0)
\end{equation}
to the leading order.  In other words, $S(t|0) \approx S_0(t|0)$ at
very short times.  However, at moderately short times, the argument of
$\erfcx(z)$ in the first term in Eq. (\ref{eq:u_asympt_moderate}) is
not necessary large.  In fact, this term vanishes at $t_* = L/(2\qc
D)$ that gives a natural timescale for the analysis of this
approximation.  Note that the argument of $\erfcx(z)$ is positive for
$t < t_*$ and negative for $t > t_*$.  Moreover, if $t \gg t_*$, one
can use $\erfcx(-z) \approx 2e^{z^2}$ for large $z$, so that
$\bar{S}(t) \sim 4 e^{-\qc L + q_{\rm c}^2 Dt}$, which is
exponentially large and thus is not applicable.

Let us summarize the above analysis.  When $\qc L$ takes small or
moderate values, the short-time behavior of $\bar{S}(t)$ is
essentially inherited from $1-S_0(t|0)$ at $t \ll t_L = L^2/(4D)$, see
Eq. (\ref{eq:u_asympt_small}).  When $\qc L$ increases, the second
timescale $t_* \ll t_L$ emerges.  If $t \ll t_*$, the leading-order
term of Eq. (\ref{eq:u_asympt_moderate}) gives precisely
Eq. (\ref{eq:u_asympt_small}), as expected.  However, the asymptotic
values in this regime are extremely small (and thus of limited
practical interest).  In turn, if $t_* \ll t \ll t_L$,
Eq. (\ref{eq:u_asympt_moderate}) is not applicable, even though this
regime is still formally qualified as ``short times''.  It simply
means that the quadratic term $\bar{S}^2(t)$ cannot be neglected in
this regime.  A thorough asymptotic analysis at large $\qc$ presents
an interesting open problem.

\end{document}